\documentclass[aps,pra,twocolumn]{revtex4}
    \usepackage{graphicx,amssymb,amsthm,bbm}

\def\ptrace#1{\mathrm{Tr}_{\mathrm{#1}}}

\def\ket#1{|#1\rangle}
\def\bra#1{\langle #1|}
\def\dens{\rho}

\def\fullDens#1{\dens_{#1}}

\def\eEntropy#1{E(\ket{#1},\{A,B\})}

\def\eDistill#1{E_D(#1,\{A,B\})}
\def\eNegativity#1{E_{\mathcal{N}}(#1,\{A,B\})}
\def\negativity#1{\mathcal{N}(#1,\{A,B\})}
\def\eFormation#1{E_{F}(#1,\{A,B\})}

\def\S#1{S_{#1}}
\def\T#1#2{T_{#1-#2,#2}}
\def\C#1#2{C_{#1,#2}}
\def\fullKet#1#2{|#2,#1\rangle}
\def\fullBra#1#2{\langle #2, #1 |}

\def\symsym#1{\widetilde{#1}}
\def\symKet#1{|\symsym{#1}\rangle}

\def\symBra#1{\langle\symsym{#1}|}
\def\symDens#1{\symsym{\rho}_{#1}}

\def\cState#1{|C(#1) \rangle}

\def\symInnerProd#1#2{\langle\symsym{#1}|\symsym{#2}\rangle}

\def\symPT#1#2{\symDens{#1}^{T_#2}}
\def\fullPT#1#2{\fullDens{#1}^{T_#2}}
\def\choose#1#2{\left(\begin{array}{c}#1\\#2\end{array}\right)}

\newtheorem{theorem}{Proposition}

\def\trace{\mathrm{Tr}}
\def\ptrace#1{\trace_{#1}}

\def\refeqn#1{Eq.\ (\ref{Equation::#1})}
\def\refeqns#1#2{Eqs.\ (\ref{Equation::#1}) and (\ref{Equation::#2})}

\def\refsection#1{Section \ref{Section::#1}}

\begin{document}

    \title{Characterizing the entanglement of symmetric many-particle spin-1/2 systems}
    \author{John K. Stockton}
    \email{jks@Caltech.EDU}
    \author{JM Geremia}
    \email{jgeremia@Caltech.EDU}
    \author{Andrew C. Doherty}
    \author{Hideo Mabuchi}
    \affiliation{Norman Bridge Laboratory of Physics 12-33,
        California Institute of Technology,
        Pasadena, California 91125 USA}

    \date{\today}
\begin{abstract}

Analyzing the properties of entanglement in many-particle spin-1/2
systems is generally difficult because the system's Hilbert space
grows exponentially with the number of constituent particles, $N$.
Fortunately, it is still possible to investigate many-particle
entanglement when the state of the system possesses sufficient
symmetry. In this paper, we present a practical method for
efficiently computing various bipartite entanglement measures for
states in the symmetric subspace and perform these calculations
for $N\sim 10^3$.  By considering all possible bipartite splits,
we construct a picture of the multiscale entanglement in large
symmetric systems.  In particular, we characterize dynamically
generated spin-squeezed states by comparing them to known
reference states (e.g., GHZ and Dicke states) and new families of
states with near-maximal bipartite entropy. We quantify the
trade-off between the degree of entanglement and its robustness to
particle loss, emphasizing that substantial entanglement need not
be fragile.

\end{abstract}

    \maketitle

\section{Introduction} \label{Section::Introduction}

The structure of entanglement within multipartite quantum systems
is a deep subject that has only begun to be explored. Because an
ensemble's Hilbert space grows exponentially with the number of
particles that comprise it, the number of distinct ways these
particles can become entangled and the number of reference states
needed to represent the various entanglement structures are
immense \cite{dur2}. While exponential scaling in complexity is
the reason multipartite entanglement is so rich, it is also the
reason the subject is so daunting.

Nonetheless, there is motivation for characterizing entanglement
in many-particle systems such as atomic spin ensembles because of
recent experimental progress in creating and manipulating
macroscopic quantum states.  In particular, highly correlated
atomic ensembles, such as spin-squeezed states \cite{kitagawa1},
have been demonstrated \cite{kuzmich, polzik1, wineland2} and
promise advances in atomic interferometry \cite{wineland1} and
quantum communication \cite{polzik2}. They also provide
experimentally accessible systems for studying quantum
measurement, feedback, and control \cite{thomsen}.

Spin-squeezing is intimately linked to the structure of the
entanglement between individual members of the ensemble
\cite{kitagawa2,wang}.  However, without a complete microscopic
picture of this entanglement, only limited claims about the
structure of these correlated states can be made. In certain
cases, an $N$-spin system can be characterized as either entangled
or separable by measuring (computing) expectation values of total
ensemble operators \cite{sorensen1,sorensen2}. For example, if the
spin-squeezing parameter for a $N$-spin state (with polarization
along $z$ and minimal variance along $x$) is less than unity
\begin{equation}
\label{Equation::sscondition}
 \frac{N \langle J_{x}^{2} \rangle}{\langle J_{z}\rangle^{2}} < 1  \\
\end{equation}
then the state is guaranteed to be inseparable. However, at this
level, limited detailed information about internal entanglement
and its robustness to particle loss \cite{dur1,rajagopal}, or other types of
decoherence \cite{kempe,andre,kitagawa2}, is available. In other
words, entanglement tests using total ensemble operators cannot
completely characterize the trade-off between the available
entanglement resources and the state's fragility.

Unlike several multipartite techniques that have been introduced
(e.g., the \textit{N-tangle} \cite{wong}), we approach the problem
of analyzing $N$-particle entanglement using only bipartite
measures.  Although a single bipartite split of a large system is
rarely sufficient to characterize multi-particle entanglement,
combining the results from many different splits of the system
paints a reconstructed picture of the many-particle entanglement.
Furthermore, by repeating the analysis after removing particles
from the system, it possible to systematically characterize the
entanglement across all size scales and its robustness to particle
loss.  Our approach has the advantage that it relies upon
well-defined entanglement measures that are both computable and
physically motivated.

Since substantial insight, and often a good starting point for
more rigorous analysis, can be gained from numerical simulations,
an efficient means of calculating entanglement measures is
desirable. \refsection{SymmetricSubspace} develops the necessary
machinery for calculating these measures in the symmetric
subspace--- the set of those $N$-particle pure states that remain
unchanged by permutations of individual particles
\cite{werner,wang,gisin}. The main result of this section is that
it is possible to perform partial transposes, partial traces, and
Schmidt decompositions of symmetric states without resorting to an
exponentially large representation of the system.

In \refsection{RepresentativeEntanglement}, we characterize
microscopic entanglement and its robustness to particle loss for
several representative symmetric states, including the GHZ and
Dicke (e.g., W) states.  Here, the advantage of exploiting
symmetry is clear; we perform entanglement calculations for
systems with $N \sim 10^3$. These numerical results allow us to
speculate on the large $N$ asymptotic scaling of the above
entanglement measures. In some cases, particularly for the
entanglement of formation and the reduced entropy, we analytically
verify the observed scaling.  We also introduce a family of states
that provides insight into the scaling of bipartite entanglement
in symmetric states for large $N$.

With the context provided by the reference states and the
boundaries of allowed entanglement structures, we can better
understand the entanglement generation abilities of certain
dynamical processes. \refsection{DynamicEntanglement} focuses on
the entanglement produced by spin squeezing Hamiltonians. We
illustrate the intuitive and generic effect that small scale
correlations peak before (and transform into) larger scale
correlations. Again, the ability to simulate systems with $N \gg
1$ permits us to determine asymptotic behavior, both for large
numbers of particles and for long times.

A point we stress is that \textit{significantly entangled states
need not be fragile}.  Robustness is critically important in
experiments where the system constantly exchanges atoms with the
surrounding environment.  Moreover, we show that spin squeezed
states provide a reasonable compromise in this trade-off; they are
highly entangled, yet particularly robust.

\section{Entanglement Measures}
\label{Section::EntanglementMeasures}

In this section, we review several common entanglement measures as
motivation for the symmetric state techniques that are developed
in Section \ref{Section::SymmetricSubspace}.  In addition to
recognizing the specific operations necessary for computing these
entanglement quantities, we also describe their strengths,
weaknesses, and, where possible, physical motivation.

We begin by reviewing the commonly accepted set of properties that
all measures of entanglement should share. For a general density
matrix, $\rho$, which can be divided into two or more sub-systems,
the quantity $E_{X}(\rho)$ (the label $X$ is used to denote a
generic measure) qualifies as an \textit{entanglement monotone} if
it satisfies the conditions \cite{vidal3,horodecki2,wei},
\begin{enumerate}

\item[(C1)]
    $E_{X}(\rho) \geq 0$;
    $E_{X}(\rho) = 0$ if $\rho$ is separable;
    $E_{X}(\text{Bell State}) = 1$.

\item[(C2)]
    Local operations, classical communication and
    post-selection (LOCC) do not increase $E_{X}(\rho)$ on
    average. For example, with any state, $\dens$, and partition, $\{A,B\}$,
    local unitary transformations, $\hat{U} = \hat{U}_A \otimes \hat{U}_B$,
    do not affect $E_{X}(\rho)$.

\item[(C3)] Entanglement is convex under discarding
    information,
    $\sum_i p_i E_{X}(\rho_i) \geq
    E_{X}(\sum_i p_i \rho_i)$.

\end{enumerate}

We define the generalized Bell States as,
\begin{eqnarray}
\ket{\Psi^{\pm}}&=&(\ket{1_A 0_B}\pm\ket{0_A 1_B})/\sqrt{2}
\\
\ket{\Phi^{\pm}}&=&(\ket{1_A 1_B}\pm\ket{0_A 0_B})/\sqrt{2}
\label{Equation::Bell}
\end{eqnarray}
for a partition $\{A,B\}$ \footnote{We work with the convention
that for any bipartite split of a system, $\{A,B\}$, the symbol
$A$ is equivalent to a list of the particles residing on its side
of the split, and likewise for $B$.  $N_A$ and $N_B$ are the
number of particles on each side of the split; $d_A$ and $d_B$ are
the Hilbert space dimensions of each side, generally $2^{N_A}$ and
$2^{N_B}$ for non-symmetric collections of spin-1/2 particles.}.
If the subsystem $A$ has more than one spin, $1_A$ is interpreted
as $1_1 \cdots 1_{N_A}$ and similarly for $1_B$, $0_A$, and $0_B$.

\subsection{Entropy of Entanglement}
\label{Section::EntropyMeasure}

Given a pure state, $\ket{\Psi}$, and a partition for the system,
$\{A,B\}$, the entropy of entanglement is defined as,
\begin{equation}
    \eEntropy{\Psi} \equiv S(\dens_A)=S(\dens_B)
\end{equation}
where the von Neumann entropy is $S(\dens)=-\trace(\dens \log_2
\dens)$ and $\dens_A=\ptrace{B}(\ket{\Psi}\bra{\Psi})$.  Any
entropy that results from performing a partial trace on the system
must be a consequence of initial entanglement provided that the
initial state is pure. For product states,
$\ket{\Psi}=\ket{\Psi}_A \otimes \ket{\Psi}_B$, the entropy is
zero since the single eigenvalue for each of the pure states
$\dens_A$ and $\dens_B$ is one. The maximum entropy of
entanglement given a partition with dimensions, $\dim(A)=d_A$ and
$\dim(B)=d_B$, with $d_A\leq d_B$, is $\log_2(d_A)$.  A state that
achieves this maximum is,
\begin{eqnarray}
    \ket{\Psi} & = &\ket{0}_A \otimes \ket{0}_B +
    \ket{1}_A \otimes \ket{1}_B + \cdots \nonumber \\
    & & +
    \ket{d_A-1}_A \otimes \ket{d_A-1}_B \label{Equation::MaxEntropyState}
\end{eqnarray}

The entropy of entanglement has the attractive feature that it is
straightforward to compute; it requires only performing a partial
trace, $\dens_A=\ptrace{B}(\dens)$, then computing eigenvalues of
the result. The drawback of the entropy is that it only qualifies
as an entanglement monotone for initially pure states.

\subsection{Entanglement of Formation}
\label{Section::FormationMeasure}

The entanglement of formation \cite{wootters1} is defined as
\begin{equation}
    \eFormation{\dens} \equiv \min\limits_{\{p_i,\psi_i\}}
    \sum_i p_i \eEntropy{\psi_i}
\end{equation}
where the $\{p_i,\psi_i\}$ satisfy the condition that
$\dens=\sum_i p_i \ket{\psi_i}\bra{\psi_i}$. This quantity is
difficult to compute for mixed states but reduces to the entropy
of entanglement for pure states.

In the special case of a mixed state of two spin-1/2 particles,
the entanglement of formation can be computed from the
two-particle concurrence, $\mathcal{C}(\dens)$
\cite{wootters1,wootters2}. Therefore, it is generally possible to
compute the entanglement of formation between two spins $\{i,j\}$
removed from an $N$-spin state $\ket{\Psi}$. The entanglement of
formation for such a reduced system is a strong measure of the
robustness of that state's entanglement to particle loss.
Explicitly, for the two particle state $\rho=\ptrace{k\neq i, j
}{\ket{\Psi}\bra{\Psi}}$
\begin{equation}
E_{F}(\rho,\{i,j\})=h(\frac{1}{2}[1+\sqrt{1-\mathcal{C}(\rho)^{2}}])
\end{equation}
where $h(x)=-x\log_{2}(x)-(1-x)\log_{2}(1-x)$ and
\begin{equation}
\mathcal{C}(\rho)\equiv\max(0,\sqrt{\lambda_1}-\sqrt{\lambda_2}-\sqrt{\lambda_3}-\sqrt{\lambda_4})
\end{equation}
in which $\lambda_1,...,\lambda_4$ are the eigenvalues of
$\rho(\sigma_{y}\otimes\sigma_{y})\rho^{*}(\sigma_{y}\otimes\sigma_{y})$
in decreasing order and $\sigma_y$ is a Pauli spin matrix.

\subsection{Distillable Entanglement and Negativity}
\label{Section::NegativityMeasure}

Given a mixed state, $\dens$, and a partition, $\{A,B\}$, the
entanglement of distillation is defined as,
\begin{equation}
    \eDistill{\dens} \equiv
    \lim\limits_{n\rightarrow\infty}\frac{m}{n}
\end{equation}
where $m$ is the number of Bell states that can be distilled from
$n$ copies of $\dens$ via an optimal purification protocol with
LOCC \cite{bennett0,bennett1}. For simplicity, we consider only
the symmetric Bell state $\ket{\Phi^{+}}$ of \refeqn{Bell} as the
output of the distillation process throughout this paper.  This
state is also known as an EPR pair, a GHZ state, or an
$N$-particle cat state. The distillable entanglement is
effectively a conversion efficiency; however, since the
purification protocol allows auxiliary separable states to be
introduced into the original system, it is possible, on average,
to extract more than one EPR pair from an initially entangled
state.  The distillable entanglement for an EPR pair is one by
definition.

The advantages of the distillable entanglement are that it is a
monotone for mixed initial states and that it quantifies
entanglement as a practical resource.  In this sense, the
distillable entanglement has a direct physical interpretation.
Unfortunately, it is extremely difficult to compute unless the
initial state is pure, in which case it reduces to the entropy of
entanglement. The entanglement of formation is an upper bound on
the distillable entanglement (i.e., one cannot extract more EPR
pairs than the number used to form the state).

There exists another entanglement monotone, the logarithmic
negativity, which, like the entanglement of formation, provides an
upper bound on the distillable entanglement but is also
\emph{computable} for mixed states \cite{vidal2}. The logarithmic
negativity is defined as
\begin{equation}
    \eNegativity{\dens}  \equiv  \log_2 (2\negativity{\dens} +1 ) \\
\end{equation}
where $\negativity{\dens}$ is the negativity of the state,
$\dens$. The negativity is defined as the absolute sum of the
negative eigenvalues of the partial transpose with respect to $A$,
$\dens^{T_A}$. So
\begin{equation}
    \negativity{\dens} \equiv \sum_i
        \frac{|\lambda_{i}|-\lambda_{i}}{2}
\end{equation}
where $\lambda_i$ are all of the eigenvalues.

The logarithmic negativity can be directly computed from the
partial transpose. However, both the logarithmic negativity and
the distillable entanglement are zero for those entangled states
with positive partial transposes (PPT).  PPT entangled states and
perhaps some other entangled states \cite{innsbruck,ibm} have zero
distillable entanglement \cite{horodecki3}.  These states are
known as \textit{bound entangled states}.

As with all monotones, the negativity may also disagree with other
monotones, like the entanglement of formation, on which state of
two is more entangled \cite{wei}. This ordering problem is a
caveat which qualifies many statements about entanglement, and is
a reflection of the fact that any given entanglement measure
refers only to its own limited physical context.

\subsection{Schmidt Decomposition}

For a given partition, $\{A,B\}$, of the full ensemble's Hilbert
space, it is possible to decompose the state as \cite{chuang},
\begin{equation} \label{Equation::SchmidtDecomposition}
    \ket{\Psi} = \sum_{i\in A} \sum_{j \in B}
        c_{ij} \ket{i}_A \ket{j}_B
\end{equation}
where the kets, $\{\ket{i}_A,\ket{j}_B\}$, provide complete bases
for $A$ and $B$, respectively.  For separable pure states, the
matrix, $\mathbf{c}$, which is not necessarily square, is rank
one, $R(\mathbf{c})=1$. States where $R(\mathbf{c}) > 1$ are
entangled because they cannot be expressed as a single tensor
product.

Generally, the Schmidt basis is taken to be diagonal in $A$. It
can be found from the matrix elements, $c_{ij}$, by performing a
singular value decomposition of $\mathbf{c}$,
\begin{equation} \label{Equation::SchmidtSVD}
    \mathbf{c} = \mathbf{U} \Lambda \mathbf{V}^\dagger
\end{equation}
where $\Lambda$ is diagonal and the rows of $\mathbf{U}$ provide
the Schmidt basis \cite{you}.  There are $r=R(\mathbf{c})$ nonzero
elements, $\lambda_1,\ldots,\lambda_r$, along the diagonal of
$\Lambda$.

Several bipartite entanglement monotones can be defined as
functions of the Schmidt coefficients \cite{vidal4,briegel},
however we present this formalism only because the Schmidt
decomposition provides an efficient procedure for computing the
entropy of entanglement. Starting with a pure state, the reduced
entropy for the partition, $\{A,B\}$, is given by,
\begin{equation}
    \eEntropy{\Psi} = -\sum_{i=1}^r \lambda_i^2 \log_2(\lambda_i^2)
\end{equation}
where the $\lambda_i$ are the singular values from
\refeqn{SchmidtSVD}.

\section{Symmetric States} \label{Section::SymmetricSubspace}

The previous section provided motivation for computing partial
traces, partial transposes, and Schmidt decompositions.  However,
for arbitrary $N$-particle spin-1/2 ensembles, these operations
are exponentially difficult to compute because a general state of
the ensemble resides in the space $\mathbbm{C}_2^{\otimes N}$ and
the dimensions of the density matrix scale as $2^N\times2^N$.
Computational investigation of arbitrary ensemble entanglement is
therefore impractical for all but the smallest values of $N$.

Fortunately, a large number of experimentally relevant states
possess symmetry under particle exchange and this property allows
us to significantly reduce the computational complexity. A large
class of $N$-particle states are invariant to symmetry
transformations of the permutation group,
\begin{equation}
    \Pi_{ij} \fullDens{N} \Pi_{ij}^{\dagger} = \fullDens{N},\quad
    \forall\,\, \Pi_{ij}
    \label{Equation::doublesideP}
\end{equation}
where the $\Pi_{ij}$ are operators that exchange particles $i$ and
$j$ within the ensemble.  This is the most general class of states
that are exchange invariant;  however, it is also possible to
further restrict the space of accessible states to those that are
symmetric with respect to single-sided permutations,
\begin{equation}
    \Pi_{ij} \fullDens{N} = \fullDens{N},\quad \forall\,\, \Pi_{ij}
    \label{Equation::singlesideP}
\end{equation}
of the individual spins.  This symmetry further constrains the
diagonal terms of the density matrix.  For the example of a two
spin system, single-sided symmetry requires
$\bra{01}\rho\ket{01}=\bra{10}\rho\ket{01}$, while the more
general double-sided symmetry does not.

The states, $\fullKet{N}{m}$, that respect this single-sided
permutation symmetry compose the symmetric subspace,
$\mathbb{S}_N$.  The ket, $\fullKet{N}{m}$, is defined as the
unnormalized $N$-particle symmetric state with $m$ excitations
(spins up),
\begin{equation}  \label{Equation::mN}
    \fullKet{N}{m} \equiv \sum_{i} P_{i}
        \left( \ket{1_1,1_2,\ldots,1_m,0_{m+1},\ldots,0_{N}} \right)
\end{equation}
where $\{P_i\}$ is the set of all $\choose{N}{m}$ distinct
permutations of the spins. Although each $\fullKet{N}{m}$ is an
element of $\mathbbm{C}_2^{\otimes N}$, the permutation symmetry
enables it to be expressed as an element, $\symKet{m}$, of a
space, $\mathbb{S}_N$, that scales linearly, rather than
exponentially, with the number of particles. In short, all states
in $\mathbb{S}_N$ can be represented in $\mathbbm{C}_{N+1}$.

The symmetric subspace therefore provides a convenient, albeit
idealized, computationally accessible class of spin states
relevant to many experimental situations (such as spin squeezing).
Completely symmetric systems are experimentally interesting
largely because it is often easier to non-selectively address an
entire ensemble of particles rather than individually address each
member.  Of course, there are still technical challenges in
preserving \emph{perfect} symmetry among the particles in an
ensemble, such as maintaining the uniformity of magnetic and
optical fields. Still, for a system of many particles,
symmetrically manipulating the ensemble generally requires fewer
resources than addressing individual members.

It is therefore attractive to consider computing various measures
of entanglement and simulating the system's dynamics using
symmetric states. However, analyzing entanglement requires at
least the operations of partial traces and partial transposes. In
order for these operations to be practical for large $N$, it is
essential to compute them in an efficient manner, i.e., without
having to work with representations of states in the full space,
$\mathbbm{C}_2^{\otimes N}$.

In this section we derive relationships that allow us to work with
arbitrary bipartite splits of the symmetric subspace. The ability
to express a symmetric state in terms of tensor products of
smaller symmetric states is a critical prerequisite for
efficiently computing bipartite entanglement measures.  In
\refsection{BasisChange} we derive the necessary expressions for
expressing symmetric states in reduced dimensional bases.  These
results lead to the operations of partial traces, partial
transposes and Schmidt decompositions on symmetric states. In all
of these cases, it is possible to manipulate symmetric states with
at worst polynomial scaling of the required computational
resources.

\subsection{Symmetric Change of Basis and Decomposition Operators}
\label{Section::BasisChange}

When working with the symmetric subspace, it is necessary to
convert between the large, $\mathbbm{C}_2^{\otimes N}$, and small,
$\mathbbm{C}_{N+1}$, basis representations of the state.  In order
to provide a systematic means for changing bases, it is convenient
to define a symmetry operator, $\S{N}: \mathbbm{C}_{2}^{\otimes
N}\rightarrow \mathbbm{C}_{N+1}$, whose action on the density
operator in the $2^N$ dimensional basis,
\begin{equation}
    \symDens{N} = \S{N} \fullDens{N} \S{N}^\dagger
\end{equation}
projects the state into $\mathbb{S}_N$ expressed in an $(N+1)$
dimensional basis.  We have adopted the notation that
$\symDens{N}$ is the symmetric density matrix represented in
$\mathbbm{C}_{N+1}$.

$\S{N}$ is an $(N+1) \times 2^N$ dimensional matrix that can be
expressed as,
\begin{equation} \label{Equation::symOp}
    \S{N} = \sum_{m=0}^N \C{N}{m} \symKet{m}\fullBra{N}{m}
\end{equation}
where the coefficients are given by,
\begin{equation}
    \C{N}{m} = \choose{N}{m}^{-\frac{1}{2}} =
    \left[\frac{N!}{m!(N-m)!}\right]^{-\frac{1}{2}}
\end{equation}
and $\C{N}{m}\fullKet{N}{m}$ is the normalized version of
\refeqn{mN}. The state $\symKet{m}$ is physically the same as the
$2^N$ dimensional state $\fullKet{N}{m}$ (both have $m$ spins up),
except that $\symKet{m}$ is normalized and expressed in the
$(N+1)$ dimensional basis,
\begin{eqnarray}
    \symInnerProd{m}{n}  & = & \delta_{m,n} \\
    \S{N} \C{N}{m} \fullKet{N}{m} & = & \symKet{m}.
\end{eqnarray}

It should be noted that $\S{N}$ is not a permutation operator, but
rather a projector.  Therefore, it is only appropriate to operate
on symmetric states with $\S{N}$ as,
\begin{eqnarray}
    \S{N} \S{N}^\dagger & = & \mathbbm{1}_{\text{sym}} \\
    \S{N}^\dagger \S{N} & \neq & \mathbbm{1}_{\text{full}}
\end{eqnarray}
where $\mathbbm{1}_{\text{sym}}$ is the identity in the $(N+1)$
dimensional symmetric basis and $\mathbbm{1}_{\text{full}}$ is the
identity in the $2^N$ dimensional full basis. Consequently,
$\S{N}^\dagger \S{N} \fullDens{N} \S{N}^\dagger \S{N} =
\fullDens{N}$ only if $\fullDens{N}$ is symmetric.  Acting on a
non-symmetric state with $\S{N}$ and $\S{N}^\dagger$ results in a
loss of information, as the non-symmetric components of that state
are lost in the projection onto $\mathbb{S}_N$.

For the purpose of making a bipartite split, $\{A,B\}$, the
essential property of the symmetric subspace is that it can be
expressed as a tensor product of smaller symmetric spaces.
However, the tensor product of arbitrary symmetric states is not
necessarily symmetric,
\begin{equation} \label{Equation::SymmetricSubset}
    \mathbb{S}_N\subset \mathbb{S}_{N-k} \otimes \mathbb{S}_k
\end{equation}
where the partition $\{A,B\}$ has been denoted by the number of
spins in each subsystem, $\{N-k, k\}$.  $\mathbb{S}_{N-k} \otimes
\mathbb{S}_k$ is larger than $\mathbb{S}_N$.  The structure of
valid symmetric products is given by the relation \cite{gisin},
\begin{equation} \label{Equation::SymmetricDecomposition}
    \fullKet{N}{m} = \sum_{p=0}^k \fullKet{N-k}{m-p}
        \otimes \fullKet{k}{p}
\end{equation}
in terms of constituent symmetric states expressed in the large
basis.

\refeqns{SymmetricSubset}{SymmetricDecomposition} raise the point
that the $N$ particle symmetric space, $\mathbb{S}_N$, is smaller
than the product space, $\mathbb{S}_{N-k} \otimes \mathbb{S}_k$.
Therefore, the entanglement of states in $\mathbb{S}_N$ will
generally be more restricted than those in the tensor product
space.  While, it is straightforward to identify the maximal
entanglement bounds for states in, $\mathbb{S}_{N-k} \otimes
\mathbb{S}_k$, the same is not true for $\mathbb{S}_N$. Therefore,
it is convenient to use the product space entanglement bounds as
an upper limit, \textit{albeit an overestimate}, for the scaling
of states in $\mathbb{S}_N$.

In order to exploit the tensor product structure in
\refeqn{SymmetricDecomposition}, motivated by our desire to
consider bipartite entanglement measures, it is beneficial to
construct a new symmetry operator, $\T{N}{k}$, that maps symmetric
states into the tensor product structure imposed by the partition,
$\{N-k,k\}$. In order to be useful for computations, both
$\mathbb{S}_{N-k}$ and $\mathbb{S}_k$ must be expressed in their
respective small bases. That is, we require the mapping $\T{N}{k}
: \mathbb{C}_{N+1} \rightarrow \mathbb{C}_{N-k+1} \otimes
\mathbb{C}_{k+1}$.

Constructing the operator, $\T{N}{k}$, can be accomplished by
decomposing $\S{N}$ according to \refeqn{SymmetricDecomposition},
\begin{equation}
    \S{N} = \sum_{q=0}^N
        \C{N}{q} \symKet{q}\left[
        \sum_{p=0}^{k} \fullBra{N-k}{q-p} \otimes \fullBra{k}{p}
        \right]
\end{equation}
and then operating on the expanded $\S{N}^\dagger$ with both
$\S{N-k}$ and $\S{k}$,
\begin{equation} \label{Equation::TOperator}
    \T{N}{k} = \sum_{q=0}^N \sum_{p=0}^{\min(q,k)}
        \frac{\C{N}{q}}{\C{N-k}{q-p}\C{k}{p}}
        \symKet{q-p}_{N-k}\otimes \symKet{p}_{k}
        \symBra{q}
\end{equation}
to produce the necessary mapping.  Here, $\symKet{m}_{N-k} \in
\mathbb{C}_{N-k+1}$ denotes symmetric states in the subsystem,
$A$, and the $\symKet{n}_k \in \mathbb{C}_{k+1}$ are symmetric
states in $B$. \refeqn{TOperator} has the interpretation of taking
an $\symKet{m} \in \mathbb{S}_N$, changing back to the large
basis, extracting the tensor product structure, and then reducing
the dimensions of the subsystems down to their respective small
bases.

\subsection{Partial Traces in the Symmetric Subspace}

In this section we derive an expression for,
\begin{equation}
    \symDens{N-k} = \ptrace{k} [ \symDens{N} ]
\end{equation}
that avoids expressing any of the density matrices (in any
intermediate step) in their large bases. The structure of the
operator, $\T{N}{k}$, immediately indicates that this is possible
since symmetric states can be expressed as tensor products of
lower-dimensional symmetric states. Once the symmetric system has
been partitioned, the partial trace is immediate.

Although the operator, $\T{N}{k}$, can be directly applied to
$\symDens{N}$, this approach condenses several intermediate steps
that might be useful when performing calculations.  Instead, we
first convert $\symDens{N}$ back to the large basis,
\begin{eqnarray}
    \fullDens{N} & = & \S{N}^\dagger \symDens{N} \S{N} \\
    \fullDens{N} & = & \sum_{m,n=0}^N \frac{ \fullBra{N}{m}
    \fullDens{N} \fullKet{N}{n} \fullKet{N}{m} \fullBra{N}{n}}{\C{N}{m}^{-2} \C{N}{n}^{-2}}
\end{eqnarray}
and then partition the symmetric states, $\fullKet{N}{m}$ and
$\fullKet{N}{n}$, using \refeqn{SymmetricDecomposition} with
$k=1$.  Taking the partial trace of the resulting expression leads
to an $N-1$ particle symmetric state in the large basis,
\begin{eqnarray}
    \ptrace{1}[\fullDens{N}] & = & \sum_{m,n=1}^{N}
     \C{N}{m}\C{N}{n}
    \fullBra{N}{m}\fullDens{N}\fullKet{N}{n} \nonumber \\
    & & \times \left[ \fullKet{N-1}{m}\fullBra{N-1}{n} \right. \nonumber\\
    & & + \left. \fullKet{N-1}{m-1}\fullBra{N-1}{n-1}\right]
\end{eqnarray}
which can be changed to the small basis using the operators,
$\S{N-1}$ and $\S{N-1}^\dagger$,
\begin{eqnarray}
    \symBra{a} \symDens{N-1} \symKet{b}
    & = &
    \C{N-1}{a}^{-1}\C{N-1}{b}^{-1}
    \left[ \symBra{a} \symDens{N} \symKet{b}
    \C{N}{a}\C{N}{b} \right. \nonumber \\
    &  & + \left.
    \symBra{a+1}\symDens{N} \symKet{b+1}
    \C{N}{a+1}\C{N}{b+1} \right]
\end{eqnarray}
By induction, it can be shown that the result of tracing $k$
particles out of the system is,
\begin{equation}
    \symBra{a} \symDens{N-k} \symKet{b} =
   \sum_{j=0}^k \symBra{a+j} \symDens{N} \symKet{b+j}
    \C{k}{j}^{-2} \frac{\C{N}{a+j}\C{N}{b+j}}{\C{N-k}{a}\C{N-k}{b}}
\end{equation}
which resides within $\mathbb{C}_{N-k+1}$.

\subsection{Partial Transposes in the Symmetric Subspace}
\label{Section::SymmetricPartialTranspose}

The structure of $\T{N}{k}$ demonstrates that the partial
transpose of symmetric states with respect to $k$ particles,
$\symPT{N}{k}$, resides in the space $\mathbb{S}_{N-k} \otimes
\mathbb{S}_k^T$, but not $\mathbb{S}_N$. Therefore, the partial
transpose involves matrices that belong to
$\mathbb{C}_{k+1}\otimes\mathbb{C}_{N-k+1}$, and computing
$\symPT{N}{k}$ scales quadratically in $N$.

As with the partial trace, the operator $\T{N}{k}$ can be directly
employed to obtain the partial transpose; however, this approach
hides several useful intermediate steps.  Instead, a more explicit
derivation involves transforming $\symDens{N}$ back to the big
basis and employing \refeqn{SymmetricDecomposition}.  The partial
transpose,
\begin{eqnarray}
    \fullPT{N}{k} & = & \sum_{m,n=0}^N \sum_{p,q=0}^k
    \C{N}{m}\C{N}{n} \symBra{m} \symDens{N} \symKet{n} \left[ \right. \\
    & & \left. \fullKet{N-k}{m-p}\fullBra{N-k}{n-q}
    \otimes\fullKet{k}{q}\fullBra{k}{p} \right] \nonumber
\end{eqnarray}
can be expressed as a tensor product,
\begin{equation}
    \fullPT{N}{k} \equiv \sum_{p,q=0}^k
    A_{N-k}^{p,q} \otimes B_{k}^{p,q}
\end{equation}
where
\begin{eqnarray}
    A_{N-k}^{p,q} & = & \sum_{m,n=0}^N\C{N}{m}\C{N}{n}
    \symBra{m} \symDens{N} \symKet{n} \nonumber \\
    & & \times \fullKet{N-k}{m-p}\fullBra{N-k}{n-q}
\end{eqnarray}
and
\begin{equation}
    B_{k}^{p,q} = \fullKet{k}{q}\fullBra{k}{p}
\end{equation}
Returning to the small basis is accomplished by evaluating,
$\tilde{A}_{N-k}^{p,q} = \S{N-k} A_{N-k}^{p,q} \S{N-k}^\dagger$
and $\tilde{B}_{k}^{p,q} = \S{k} B_{k}^{p,q} \S{k}^\dagger$ to
give,
\begin{eqnarray}
    \symBra{a}  \tilde{A}_{N-k}^{p,q} \symKet{b} & = &
    \frac{\C{N}{p+a}\C{N}{q+b}}{\C{N-k}{a}\C{N-k}{b}}
    \symBra{a+p}\symDens{N} \symKet{b+q} \\
    \symBra{c} \tilde{B}_{k}^{p,q} \symKet{d}
    & = & \C{k}{c}^{-1}\C{k}{d}^{-1} \delta_{q,c}\delta_{p,d}
\end{eqnarray}
where,
\begin{equation}
    \symPT{N}{k} = \sum_{p,q=0}^k
    \tilde{A}_{N-k}^{p,q} \otimes \tilde{B}_{k}^{p,q}
\end{equation}
shows that the dimension of $\symPT{N}{k}$ is, in fact,
$(k+1)\times(N-k+1)$.

\subsection{Schmidt Decomposition of the Symmetric Subspace}
\label{Section::SymmetricSchmidt}

It is quite simple to perform the Schmidt decomposition,
\refeqn{SchmidtDecomposition}, of a symmetric state in
$\mathbb{S}_{N}$, into the space $\mathbb{S}_{N-k} \otimes
\mathbb{S}_k$. The coefficients, $\mathbf{c}$, in
\refeqn{SchmidtDecomposition} for the states $\symKet{m}$ follow
directly from applying the operator $\T{N}{k}$ to $\symKet{m}$,
resulting in the expression,
\begin{equation} \label{Equation::SymmetricSchmidt}
    \T{N}{k}\symKet{m} = \sum_{i=0}^{N-k} \sum_{j=0}^k
    \delta_{m,i+j}\frac{\C{N}{m}}{\C{N-k}{i}\C{k}{j}}
    \symKet{i}_{N-k}
    \otimes \symKet{j}_k
\end{equation}
For the states, $\symKet{m}$, the Schmidt matrix, $\mathbf{c}$,is
sparse and the singular value decomposition, \refeqn{SchmidtSVD},
can be performed analytically.

General symmetric states, $\symKet{\Psi} = \sum_{m=0}^N a_m
\symKet{m}$, can be represented as
\begin{equation} \label{Equation::SymmetricSchmidtPsi}
    \T{N}{k}\symKet{\Psi} = \sum_{m=0}^N a_m
    \sum_{i=0}^{N-k} \sum_{j=0}^k
    \delta_{m,i+j}\frac{\C{N}{m}}{\C{N-k}{i}\C{k}{j}}
    \symKet{i}_{N-k}\symKet{j}_k
\end{equation}
However, for these general symmetric states, the Schmidt
coefficient matrix, $\mathbf{c}$, is not sparse.

\subsection{Dynamics in the Symmetric Space}

One of the objectives of this paper is to treat dynamically
generated entangled states, therefore this section briefly
discusses the time evolution of symmetric states.   It is
straightforward to show that acting on a symmetric state with
operators of the form,
\begin{equation} \label{Equation::SymOperator}
    \mathbf{o} = \sum_{j=1}^N \mathbbm{1}^{(1)}
    \otimes \cdots \mathbf{o}^{(j)}
    \cdots \otimes \mathbbm{1}^{(N)}
\end{equation}
preserves the exchange symmetry in the large basis, $[\mathbf{o},
\Pi_{ij}]=0$, provided that the $\mathbf{o}^{(i)}$ are identical.

Using the symmetric state change of basis operator, $\S{N}$,
elucidates the physical nature of the symmetric subspace. For
example, transforming any angular momentum operator of the form in
\refeqn{SymOperator} to the small basis using $\S{N}$
\begin{equation}
    \tilde{\mathbf{J}} = \S{N}\mathbf{J}\S{N}^\dagger
    =  \S{N}\left(\sum_i \mathbf{j}^{(i)}\right)\S{N}^\dagger
\end{equation}
produces the $(N+1)$ dimensional operator equivalent to the
angular momentum for a single \emph{pseudo-spin} $J=N/2$ particle.
This is because the symmetric subspace is composed of basis states
$\symKet{m}$ that correspond to the eigenstates of $J_z$ with
$J=N/2$ (e.g., for two spins, the symmetric subspace includes the
triplet, but not the singlet).

The dynamics of any symmetric state are confined to the symmetric
subspace provided that the Hamiltonian can be expressed as a
function of operators all of the form \refeqn{SymOperator}. Given
a symmetry-preserving Hamiltonian, the dynamics can be completely
simulated with the small symmetric basis. Explicitly, an
infinitesimal step of evolution can be written
\begin{eqnarray}
    \ket{\tilde{\Psi}(t+dt)} & = & \S{N}(1 + i H dt)\ket{\Psi(t)} \nonumber \\
&=&\S{N}\ket{\Psi(t)}+i dt \S{N} H \ket{\Psi(t)} \nonumber \\
&=&\S{N}\ket{\Psi(t)}+i dt \S{N} H \S{N}^{\dagger} \S{N} \ket{\Psi(t)} \nonumber \\
&=&\ket{\tilde{\Psi}(t)}+i dt \tilde{H}  \ket{\tilde{\Psi}(t)}
\end{eqnarray}
where we have used $\ket{\tilde{\Psi}(t)}=\S{N} \ket{\Psi(t)}$,
$\tilde{H} \equiv \S{N} H \S{N}^{\dagger}$, and
$\ket{\Psi(t)}=\S{N}^{\dagger} \S{N} \ket{\Psi(t)}$ (because
$\ket{\Psi(t)}$ is assumed symmetric).

For many experimentally motivated $N$-particle spin-1/2 systems,
it is possible to express states using the symmetric subspace and
the dynamics using only symmetry-preserving operators. The only
time this efficient representation fails to apply is when the
symmetry is broken or the system is divided (as we consider
throughout the paper). For example, the spontaneous local decay of
any one spin is sufficient to break the symmetry of
\refeqn{singlesideP}. Depending on the form of the decoherence,
some symmetry may be retained (e.g., the particle exchange
symmetry of \refeqn{doublesideP}). Other treatments have addressed
the effect of such decoherence on parameters related to
entanglement, such as the degree of spin squeezing
\cite{kempe,kitagawa2,andre}.

\section{Entanglement Properties for Representative Symmetric States}
\label{Section::RepresentativeEntanglement}

Given the large number of possible $N$-spin states, even when
restricted to the symmetric space, it is clear that a systematic,
yet compact approach to characterizing microscopic entanglement is
necessary. Toward this end, we characterize a set of
representative symmetric states with a limited combination of
measures, including the reduced state entropy, the entanglement of
formation, and the logarithmic negativity. The families (described
in detail below) that we have selected display diverse
entanglement behavior--- they differ in their degree of
entanglement at different size scales and in their robustness to
particle loss. Naturally, any set of representative states will be
incomplete in some aspect; however, our goal is to provide a
detailed picture of internal entanglement without an excessive
number of representatives.

In this section we address the relationship between the degree of
entanglement and its robustness to particle loss. While it has
been a longstanding conception that the most entangled states are
simultaneously the most fragile, we demonstrate that this is not
necessarily true.  Under certain useful definitions of
entanglement, it is possible to find heavily entangled symmetric
states that are simultaneously robust. Similarly, the most fragile
states are not always the most entangled.  We also demonstrate
that restricting our analysis to the symmetric subspace does not
preclude the potential for significant entanglement.

\subsection{Symmetric Reference States}
\label{Section::ReferenceStates}

We now briefly describe several families of representative
symmetric states using the notation introduced in
\refsection{SymmetricSubspace}.  In addition to GHZ states and the
W family, we introduce a new parametrized family, termed ``comb
states'', which prove important in investigating the maximal
boundary of certain entanglement measures.

Throughout the rest of this paper, all states are assumed to be
symmetric.  In the interest of simpler notation, we will express
symmetric states as $\rho$ even when it is more efficient to
compute entanglement measures using their $\tilde{\rho}$
representation. Tilde notation is used only for the $\symKet{m}$
states.

\subsubsection{GHZ States} \label{Section::GHZStateDefinition}

The well known GHZ states \cite{GHZ} can be written,
\begin{equation}
    \ket{\mathrm{GHZ}} = ( \symKet{0} + \symKet{N} ) / \sqrt{2}
\end{equation}
using the notation from \refsection{SymmetricSubspace}.  The GHZ
family is generally considered to be the standard example of a
highly entangled state.  In several different contexts it has
become the common unit of entanglement currency. For example, as a
particular Bell state, the GHZ state is the end goal of
entanglement distillation protocols.

However, the GHZ family fails to maximize a number of monotones,
including the entanglements of distillation and formation for a
given bipartite split. Unlike previous treatments \cite{gisin}, we
choose to work with these measures under which the GHZ is not a
maximally entangled state. Certain other measures such as the
\emph{N-tangle} correctly recognize the GHZ as containing the most
true $N$-way entanglement \cite{wong, wootters3, briegel}, but our
focus will remain on notions of strictly bipartite entanglement.
Still, the most practical defining characteristic of the GHZ state
is its fragility to particle loss; tracing out a single party
destroys \textit{all} of the internal entanglement.

\subsubsection{Dicke States} \label{Section::WStateDefinition}

An important family of states with completely different character
is the set of symmetric states with integer $m$ excitations (spins
up), $\symKet{m}$, where $m=0,\ldots,N$.  Of course, these states
are also known as the Dicke states or the eigenstates of $J_z$,
where the notation $|J,M\rangle$ is used with $J=N/2$ and
$M=m-N/2=-N/2,\ldots,N/2$. The W state \cite{dur3}, which is
defined as the symmetric state with one excitation,
$\ket{\mathrm{W}} \equiv \symKet{1}$, is a particular member of
this family. Notice that $\symKet{m}$ and $\symKet{N-m}$ have the
same entanglement properties because one is equal to the other if
the quantization axis is reversed. These states exhibit a high
degree of entanglement for $m=1,\ldots,N-1$, while the states of
$m=0,N$ are completely separable. The defining characteristic of
the Dicke state entanglement is remarkable robustness to particle
loss. It has been proven that $\symKet{1}$ optimizes the
concurrence when all but two spins have been removed
\cite{koashi}, the extreme opposite of the fragile GHZ behavior.
It has also been proven that for single copies the GHZ and W
cannot be converted into each other with LOCC operations on the
individual spins with certainty \cite{dur3}, further emphasizing
their difference. For additional discussions of the conversion
properties of entangled states see
\cite{verstraete,bennett2,vidal5}.

\subsubsection{Comb States} \label{Section::CStateDefinition}

A parametrized family of practical importance, which we call comb
states, are defined as
\begin{equation}
\ket{\mathrm{C}(s)} = \sqrt{\frac{2s}{N}}
        \sum_{m=-N/s}^{N/s} \symKet{N/2+ms}
\end{equation}
In the $\symKet{m}$ basis, these states have a comb-like structure
with $m$-independent weighting for the non-zero elements which are
spaced by $s$ excitations.  Because the comb states non-trivially
explore the full support of the symmetric basis they may be
expected to access regions of entanglement space where
$\symKet{m}$ states are forbidden.  Luckily, particular comb
states with an optimized spacing $s$ turn out to contain near
maximal entanglement for bipartite splits of any symmetric
ensemble as will be shown numerically and proven in the appendix.

\subsubsection{Random States}

Another way to numerically explore the full symmetric space is
randomly. We define a randomly generated state
$\ket{\mathrm{R}}=\sum_m r_m \symKet{m}$ where the coefficients
$r_m$ are complex Gaussian random variables with averages
$\mathrm{E}[r_m]=0$, $\mathrm{E}[r_m r_n]=0$, and
$\mathrm{E}[r_m^*r_n]=\delta_{mn}/(N+1)$. Note that this
distribution of states is independent of the basis in terms of
which we have chosen to define the random coefficients $r_m$. If
we write $\ket{\mathrm{R}}=\sum_m r'_m (U|\tilde{m}\rangle)$ in a
new basis $U|\tilde{m}\rangle$, where $U$ is an arbitrary unitary,
the new coefficients $r'_m$ have exactly the same Gaussian
distribution as the coefficients $r_m$. As a result this
distribution determines a measure on (unnormalized) vectors in
$\mathbb{C}_{N+1}$ that is invariant under unitary
transformations. Moreover $\mathrm{E}[\langle \mathrm{R} |
\mathrm{R} \rangle]=1$ so the states are on average normalized and
in fact the distribution of norms becomes very sharply peaked
around one as $N\rightarrow \infty$. In this limit we can regard
the states $\ket{\mathrm{R}}$ as being drawn from the natural
unbiased distribution of pure states. In practice we randomly
select these vectors for a fixed finite $N$ of interest and
normalize.

\subsection{Pure State Entropy of Entanglement}
\label{Section::SymmetricEntropy}

For an initially pure, $N$-particle symmetric state, there are
$\lfloor N/2 \rfloor$ possible ways to partition the system into
two parts. With symmetric states we can replace the labelling of a
particular partition $\{A,B\}$ with the number of spins in each
partition $\{N_A = k, N_B = N-k\}$ where $k = 1,\ldots,\lfloor N/2
\rfloor$. The entropy can then be computed from either of the
reduced density matrices \cite{wootters1}, $\rho_{N-k} =
\ptrace{k}\rho_{N}$ or $\rho_{k} = \ptrace{N-k} \rho_{N}$:
\begin{eqnarray}
    E({\ket{\Psi}},\{k,N-k\})
        &=& S(\ptrace{k}\ket{\Psi}\bra{\Psi}) \nonumber \\
        &=& S(\ptrace{N-k}\ket{\Psi}\bra{\Psi})
\end{eqnarray}
It can be proved that the entropy is a monotonically increasing,
concave down function of $k$ in this range \cite{ohya}. (From this
point on, we implicitly assume the rounding of non-integer numbers
such that $\lfloor N/2 \rfloor$ is implied by $N/2$ and $\{\lfloor
N/2 \rfloor,\lceil N/2 \rceil\}$ is implied by $\{N/2, N/2\}$.)

In \refsection{BasisChange}, we emphasized that a symmetric state
with $N$ particles can be represented on the product space of two
symmetric spaces with $N-k$ and $k$ particles ($\mathbb{S}_{N-k}
\otimes \mathbb{S}_{k}$). For all states within this space, the
state of \refeqn{MaxEntropyState} (with $d_{A}=k+1$) maximizes
$E({\ket{\Psi}},\{k,N-k\})$ at $\log_{2}(k+1)$. However, this
state is \emph{not} symmetric with respect to the exchange of any
two particles across the split. We are interested in finding the
upper bound for the states in the space $\mathbb{S}_N$, which are
only a subset of states in $\mathbb{S}_{N-k} \otimes
\mathbb{S}_{k}$. It has been proven that the additional
restriction of overall symmetry constrains the maximal entropy to
be strictly less than $\log_{2}(k+1)$ except for $N=2,3,4$ and $6$
where states can be found that achieve this bound \cite{gisin}.
Consequently, we refer to the bound $\log_{2}(k+1)$ as the
\emph{unobtainable bound} for any $k$.

\begin{figure}
\centerline{\includegraphics[width=3.2in]{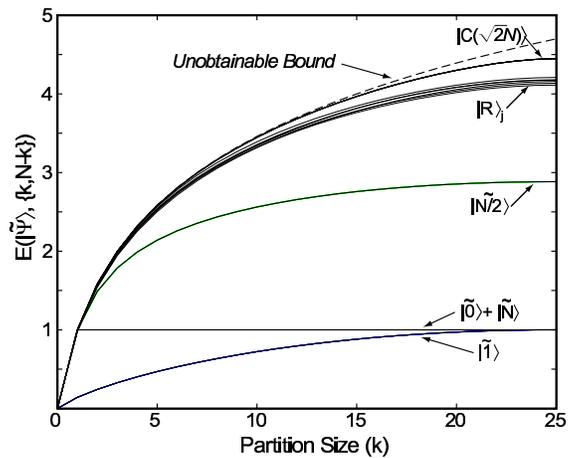}}
\caption{Entropy of entanglement for representative symmetric
states (described in \refsection{RepresentativeEntanglement}) with
$N=50$ particles as a function of the dimension of the bipartite
split, $\{k,N-k\}$, where $k = 1,\ldots,\lfloor N/2 \rfloor$.  The
\textit{unobtainable bound} $\log_2(k+1)$ is the entropy that
could be achieved by a non-symmetric product of the two symmetric
subsystems, $\{A,B\}$.  Several representative states nearly
achieve this maximum.\label{esplitsVsplitref}}
\end{figure}

Figure \ref{esplitsVsplitref} shows a plot of
$E({\ket{\Psi}},\{k,N-k\})$ for several reference states and
$N=50$. Despite the fact that all states are forbidden from
achieving the value $\log_{2}(k+1)$, some states come close to
achieving this unobtainable bound. These include most randomly
generated states and the comb states with $s=\sqrt{2N}$. This
naturally leads us to the question: what exactly is the minimum
upper bound for the split entropy of symmetric states and what
states achieve that bound?

\begin{figure*}
\centerline{\includegraphics[width=6.4in]{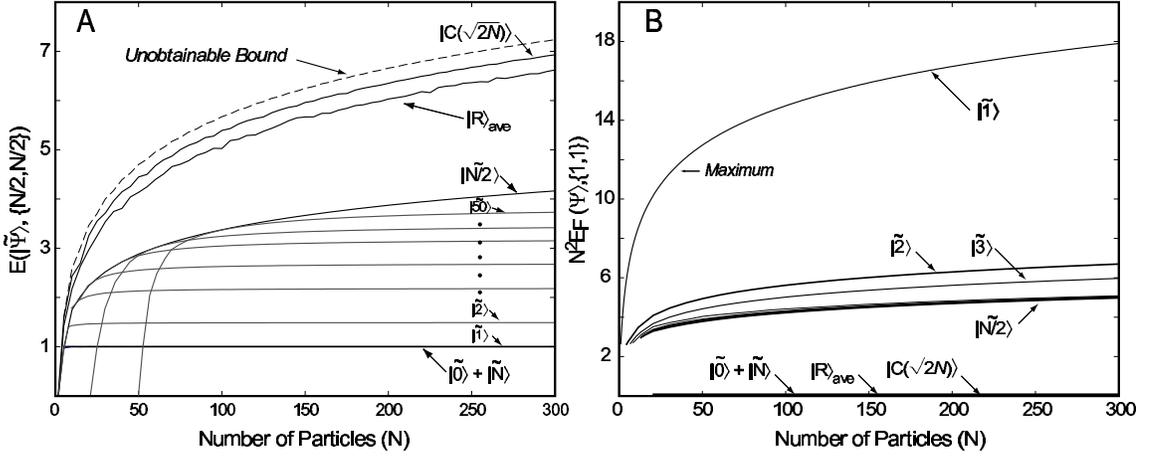}}
\caption{(A) Plot of the even split entropy of entanglement,
$E(\rho,\{N/2,N/2\})$, for representative states as a function of
the number of particles, $N$ (which is also equal to the
entanglement of formation and distillation). Note that the average
entropy of $25$ random states, $\ket{R}$, as well as the entropy
of $\cState{\sqrt{2N}}$, nearly attain the \textit{unobtainable
bound} $\log_2(\lfloor N/2 \rfloor +1)$. (B) A plot of the
two-particle entanglement of formation, $E_F(\rho,\{1,1\})$, as a
function of the number of particles, $N$. The W state,
$\symKet{1}$, maximizes this entanglement measure, which
quantifies robustness to particle loss. \label{EVNref}}
\end{figure*}

\subsubsection{Maximizing the Even Split Entropy}

Because the entropy is maximized by the most even split ($k= N/2
$), we henceforth consider only this partition.  From the above discussion,
we know that for $N \geq 7$, the entropy obeys the inequalities,
\begin{equation}
E(\ket{\Psi},\{N/2, N/2\})\leq E_{max}(N) < \log_2(N/2 + 1)
\end{equation}

Analytically locating the minimum upper bound $E_{max}(N)$ (or the
states that achieve it) is difficult, but a simple numeric
approach turns out to shed some light on what we can expect.
Figure \ref{EVNref}A shows the entropy of the even split entropy
as a function of $N$ for several families of states with the
unobtainable upper bound for reference.  Most families of states
do not keep up with the scaling of this upper bound.

For example, if $N \gg m$ the states $\symKet{m}$ (with
$m=1,\ldots,N/2 $) can be shown to have entropies of
\begin{equation}
E(\symKet{m},\{N/2, N/2 \}) \approx \frac{\log_2(m)}{2}+1
\end{equation}
We also see that the largest of these scales as
$E(\symKet{N/2},\{N/2, N/2 \}) \approx \log_2(N)/2$.

Because of the factor of two, none of these states keep up with
the $\mathbb{S}_{N/2} \otimes \mathbb{S}_{N/2}$ bound.  However,
if we explore the simplest possible states accessing more of the
symmetric Hilbert space, we find something quite different.  For
large $N$ (up to $600$), the average entropies of random states,
for example, numerically scale as $\approx \log_{2}(N/2+1)-0.6$.
This indicates the remarkable fact that the symmetry constraint on
the overall state does not limit the \emph{scaling} of the maximal
bipartite entanglement compared to that of the more general space
$\mathbb{S}_{N/2} \otimes \mathbb{S}_{N/2}$.

The comb states, optimized over the spacing $s$, are even more
entangled. Numerically, we find that (for $N$ up to $600$) their
entropies scale as $\approx \log_{2}(N/2+1)-0.3$, when
$s\approx\sqrt{2N}$. Encouraged by this evidence, we were able to
prove in the asymptotic limit of large $N$ that this family of
comb states $\cState{\sqrt{2N}}$ does indeed scale as
$\log_{2}(N/2+1)-\delta$ where $\delta$ is a constant of order
unity (see appendix). A similar proof for the random state scaling
is probably possible. The fact that random states, and the
optimized comb state, seem to nearly maximize the $\{N/2,N/2\}$
entropy indicates that the set of states which scale similarly is
of non-zero measure (i.e., this behavior is not atypical).

Still we have not located the value of the true minimum upper
bound and the form of the states that achieve this bound. Given
the above results we expect it to have a similar scaling with a
minimal offset, $\delta$, for large $N$.

\subsection{Entanglement of Formation: Extremal Splits}

For any bipartite entanglement measure, we can construct even more
possible splits if we choose (or are forced) to ignore some of the
particles. Suppose we start with a symmetric state of $N$ spins
$\ket{\Psi}$ and trace out spins until only $N_r$ remain. In this
case, the new state
$\rho_{N_r}=\ptrace{N-N_r}(\ket{\Psi}\bra{\Psi})$ will be mixed
but \emph{still} symmetric.  We then have the possible bipartite
splits $\{k,N_r-k\}$ with $k=0,\ldots,N_r/2$ \footnote{We use the
convention that for any bipartite measure X if $j+k$ is less than
the number of spins $N$ in $\ket{\Psi}$, the possibly mixed state
that the measure $E_{X}(\ket{\Psi},\{j,k\})$ should act on is
actually $\ptrace{N-j-k}\ket{\Psi}\bra{\Psi}$.  For the
concurrence we use the similar convention
$\mathcal{C}(\ket{\Psi})=\mathcal{C}(\ptrace{N-2}\ket{\Psi}\bra{\Psi}$).
\label{convention}}.

For pure states, the entropy of entanglement for any bipartite
split is equal to both the entanglement of formation and
distillation.  Unfortunately, numerically calculating either of
these monotones is much more difficult if given an initially mixed
density matrix.  For negativities, we showed in section
\ref{Section::SymmetricPartialTranspose} that we can numerically
calculate all bipartite splits $\{k,N_r-k\}$ for symmetric states,
and we will demonstrate this ability in section
\ref{Section::NegAnalysis}.  For now we would like to deal with
the extreme case of all but two spins removed ($N_r=2$). In
section \ref{Section::FormationMeasure}, we stated that the
entropy of formation $E_F(\ket{\Psi},\{1,1\})$ is easily
calculated for two spin mixed states through the concurrence. By
discussing the relationship of the pair
$[E_F(\ket{\Psi},\{1,1\}),E_F(\ket{\Psi},\{N/2,N/2\})]$, we can
start to get a sense of the allowed relationship of entanglement
across the extremes of size scales.  We will refer to the splits
$\{1,1\}$ and $\{N/2,N/2\}$ as the \emph{extremal splits}.

Figure \ref{EVNref}B displays $E_F(\ket{\Psi},\{1,1\})$ for
several reference states.  It has been proven that the W state
$\symKet{1}$ maximizes the concurrence, hence also the
entanglement of formation (for all symmetric states) with a value
of $\mathcal{C}(\symKet{1})=2/N$ \cite{koashi,dur1}. Wang and
M{\o}lmer \cite{wang} have shown that by using a similar formalism
where the two spin concurrences are calculated from the moments of
the entire state, analytic expressions can be derived for the
concurrences of several families of symmetric states.  In
particular, for the Dicke states $\symKet{m}$, and $M=m-N/2$, the
concurrence is
\begin{eqnarray}
&&\mathcal{C}(\symKet{M+N/2})=\frac{1}{2N(N-1)}\{N^2-4M^2 \nonumber\\
&&\,-\sqrt{(N^2-4M^2)[(N-2)^2-4M^2]}\}
\end{eqnarray}
which gives the above result for the $\symKet{1}$ state and also
$\mathcal{C}(\symKet{N/2})=1/(N-1)$. In the large $N$ limit these
concurrences lead to the entanglements of formation
\begin{eqnarray}
E_F(\symKet{1},\{1,1\})&\approx&\frac{2\log_2{N}+\log_2{e}}{N^2}\\
E_F(\symKet{N/2},\{1,1\})&\approx&\frac{2\log_2{2(N-1)}+\log_2{e}}{(2(N-1))^2}
\end{eqnarray}

The $1/N^2$ scaling is due to the fact that the two spin state is
constrained to be reduced from a larger symmetric $N$ spin state.
In effect, one spin can only be so entangled with another when it
is constrained to have the same relationship with all other spins.

For many states $E_F(\ket{\Psi},\{1,1\})$ is simply zero. The GHZ
state, the comb state, and practically all random states have zero
$\{1,1\}$ entanglement and do not contain the same degree of
robust entanglement as the $\symKet{m}$ states. Furthermore the
ordering of states shown in figure \ref{EVNref} A and B is
reversed (with the exception of the GHZ). This leads to the
question: what is the nature of the trade-off between the small
and large scale entanglement of the extremal splits?

\begin{figure}
\centerline{\includegraphics[width=3in]{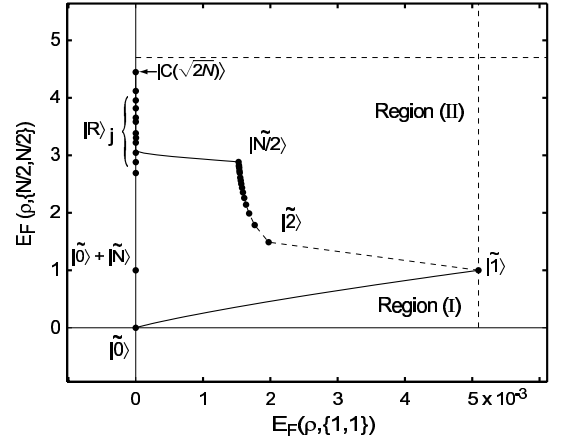}}
\caption{Plot of accessible entangled states in a space that
reflects the trade-off between the degree of entanglement
($E_F(\rho,\{N/2,N/2\})$) and its robustness to particle loss
($E_F(\rho,\{1,1\})$). The degree of entanglement for the large
split is also equal to the distillable entanglement.
\label{eNVe2ref}}
\end{figure}

Figure \ref{eNVe2ref} shows each state as a point in the space of
$[E_F(\ket{\Psi},\{1,1\}),E_F(\ket{\Psi},\{N/2,N/2\})]$ for
$N=50$. The line between $\symKet{0}$ and $\symKet{1}$ represents
states which are a linear combination of these two states.  The
curve extending from the $\symKet{N/2}$ state to the vertical axis
and up that axis to the comb state represents linear combinations
of those two states. The forbidden regions of this space for
symmetric states are unknown but we strongly suspect several
properties of the boundaries. We conjecture that there are two
regions where no states are allowed to exist. First, in region
(I), beneath the $\symKet{0}\leftrightarrow\symKet{1}$ line, no
states are found, nor likely to exist.  The reason for this is
that to have any $\{1,1\}$ entanglement there must exist some
degree of $\{N/2,N/2\}$ entanglement.  However, there must also be
a region in the upper right (II) where no states exist. It appears
that there exists a fundamental trade-off between small and large
scale entanglement: as the large scale entanglement of a state
increases (and is above $1$), the maximum allowable small scale
entanglement will decrease.  In addition, there is likely a
critical value of the $\{N/2,N/2\}$ entanglement above which the
$\{1,1\}$ entanglement must be zero.

\subsection{Negativities: Extension to All Splits}
\label{Section::NegAnalysis}

Now that we have a better intuition for the relationship between
the entanglement of the extremal splits, we can more confidently
approach the problem of understanding the large number of
remaining splits.  For $N_r$ spins remaining, there are $N_r/2$
splits of the form $\{k,N_r-k\}$ with $k=1,\ldots,N_r/2$. If
$N_r<N$ and the initial state is non-separable, the reduced state
is mixed and one of the few computable entanglement measures
available is the negativity. Even though it is a computable
monotone, the negativity is not an entanglement measure with as
much physical justification as the entanglement of formation or
distillation. However, the logarithmic negativity is an upper
bound for the distillable entanglement \cite{vidal2}. With this in
mind, we move forward and work with the logarithmic negativities
as an indicator of \emph{potential} entanglement.

\subsubsection{Negativity of All Even Splits}

Before computing the negativities, we can use the properties of
monotones to notice a few relationships between the bipartite
monotones of different splits. Tracing out a single spin is an
operation that falls under LOCC, and any monotone $X$, including
the negativity, can only decrease under such an operation,
therefore
\begin{eqnarray}
E_{X}(\ket{\Psi},\{k-1,N_r-k\}) & \leq & E_{X}(\ket{\Psi},\{k,N_r-k\}) \nonumber \\
E_{X}(\ket{\Psi},\{k,N_r-k-1\}) & \leq &
E_{X}(\ket{\Psi},\{k,N_r-k\}) \nonumber
\end{eqnarray}

\begin{figure}
\centerline{\includegraphics[width=3in]{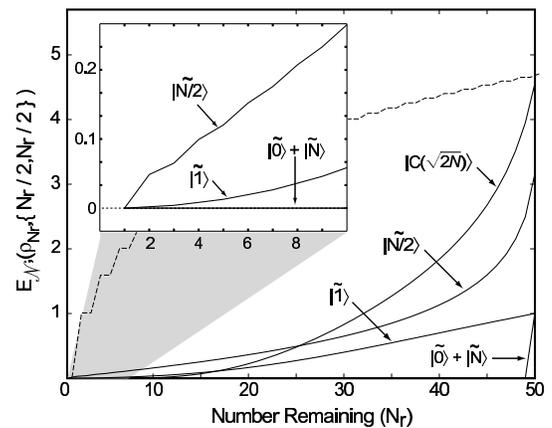}}
\caption{Plot of the even split negativity,
$E_{\mathcal{N}}(\rho,\{N_r/2,N_r/2\})$, for representative
symmetric states with $N=50$, as a function of the number of
particles remaining, $N_r$, in the system.  The inset plot
highlights the particular robustness of the $\symKet{N/2}$ Dicke
state as measured by the negativity.  This contrasts a similar
analysis using the entanglement of formation, where $\symKet{1}$
is most robust. \label{negsVsplitref}}
\end{figure}

For pure states, the most even split $\{N/2,N/2\}$ gives the
maximal entropy of entanglement \cite{vidal1}.  We observe that
this is also true for the most even splits of a reduced mixed
state with $N_r$ particles remaining $\{N_r/2,N_r/2\}$.  These
observations motivate us to reduce the number of splits considered
to only the even splits of a given $N_r$. Figure
\ref{negsVsplitref} displays the quantity
$E_{\mathcal{N}}(\ket{\Psi},\{N_r/2,N_r/2\})$ as a function of
$N_r$ for several reference states.  The endpoints of this plot
give similar information about the extremal splits as the previous
description of entanglement of formation.  Unlike the entanglement
of formation, we can easily plot the intermediate splits for the
logarithmic negativity.

By the above inequalities, we know that each curve monotonically
increases with $N_r$.  For reference we have included the plot of
$\log_{2}(N_r/2+1)$ which of course cannot be achieved, because
each reduced state with $N_r$ spins remaining is constrained by
the symmetry of the initial pure state. The space between this
maximum and the space of all actual curves represents the
entanglement `cost' of initial symmetrization. An unanswered
question is for a given $N_r$ and $N$ what pure state $\ket{\Psi}$
maximizes $E_{\mathcal{N}}(\ket{\Psi},\{N_r/2,N_r/2\})$? What is
this maximum as a function of $N_r$ and $N$?  These questions for
both the negativities and other bipartite monotones are extensions
of the problems encountered for the extremal splits.  Again, we
plot only the reference states and set aside the problem of fully
characterizing the space of interest.

First consider the GHZ state $\symKet{0}+\symKet{N}$.  As
expected, this state is maximally fragile, starting at unity and
dropping to zero as soon as one spin is removed.  In direct
contrast, the W state $\symKet{1}$ starts at unity, but only
slowly decays to zero as spins are removed and its logarithmic
negativity remains finite for even $N_r=2$.  The state
$\symKet{N/2}$ is, in some sense, an optimal trade-off between
total entanglement and robust entanglement in that it starts
reasonably high above unity at $N_r=N$, but appears to have
maximal negativity below $N_r\approx N/2$. The comb states (and
random states), which have near maximal total entanglement, are
also a reasonable trade-off, especially compared to the extreme
fragility of the GHZ state.

The comb state and most random states still `bottom out' with zero
negativity (no negative eigenvalues of the partial transpose)
below a critical $N_r$. Because the logarithmic negativity is an
upper bound on the distillable entanglement, this must also be
zero at these points. The size of this critical $N_r$ for a given
state is another indicator of fragility of the entanglement (for
the GHZ state it is the extreme $N-1$). For the optimal comb
state, the size of this critical value appears to scale only
logarithmically in $N$. Thus the comb states, despite having near
maximal $\{N/2,N/2\}$ entanglement, contain entanglement that can
withstand a huge amount of particle loss.

\subsubsection{Ordering of Dicke States}

\begin{figure}
\centerline{\includegraphics[width=3in]{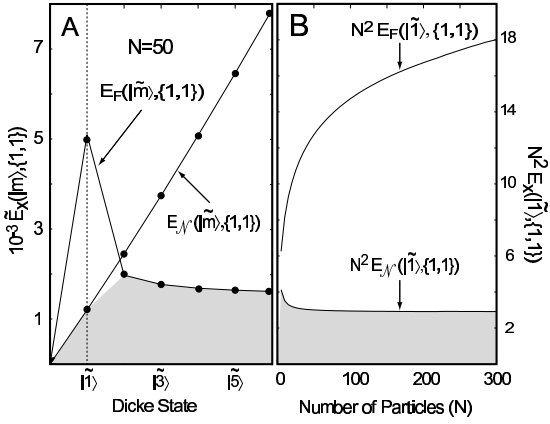}}
\caption{(A) Plot of the inconsistent ordering of the reduced
entanglement of formation, $E_F(\rho,\{1,1\})$, and the reduced
logarithmic negativity, $E_{\mathcal{N}},\{1,1\})$, for Dicke
states.  The shaded region reflects the possible values for the
distillable entanglement. (B) The large $N$ scaling of the
entanglement measures in (A) for the $\symKet{1}$ (i.e., W) state.
\label{batman}}
\end{figure}

Given the fact that $\symKet{1}$ optimizes the entanglement of
formation of the $\{1,1\}$ split, it may seem odd that
$\symKet{N/2}$ maximizes the negativity. Indeed, there is an
ordering issue here and the two monotones disagree on which of the
reduced states is more entangled.  See \cite{wei} for a more
complete discussion of ordering problems with entropies and
entanglement measures for two spin systems. Figure \ref{batman}
displays the ordering problem between
$E_{\mathcal{N}}(\symKet{m},\{1,1\})$ and
$E_{F}(\symKet{m},\{1,1\})$ for $N=50$. For $N/2\geq j \geq 1$,
$E_{F}(\symKet{j},\{1,1\})>E_{F}(\symKet{j+1},\{1,1\})$ whereas
$E_{\mathcal{N}}(\symKet{j},\{1,1\})<E_{\mathcal{N}}(\symKet{j+1},\{1,1\})$,
so the quantities are respectively decreasing and increasing with
$j$. In fact the two curves will always cross because,
$E_{F}(\symKet{1},\{1,1\}) > E_{\mathcal{N}}(\symKet{1},\{1,1\})$
and $E_{F}(\symKet{N/2},\{1,1\}) <
E_{\mathcal{N}}(\symKet{N/2},\{1,1\})$. For large $N$, $N^2
E_{F}(\symKet{N/2},\{1,1\}) \approx \log_{2}(N)/2 < N^2
E_{\mathcal{N}}(\symKet{N/2},\{1,1\}) \approx N\log_{2}(e)$ where
the approximations can be shown both analytically and numerically.

For $N \gg m$, $N^2 E_{\mathcal{N}}(\symKet{m},\{1,1\})$ flattens
out to a constant as a function of $N$, while $N^2
E_{F}(\symKet{m},\{1,1\})$ continues to grow logarithmically, as
shown if figure \ref{batman}B for $\symKet{1}$. In this case, the
entanglement of formation is significantly greater than the
logarithmic negativity and hence also the distillable
entanglement.  So, for the state $\symKet{1}$, we can show
\begin{eqnarray}
E_{D}(\symKet{1},\{1,1\}) &\leq&
E_{\mathcal{N}}(\symKet{1},\{1,1\}) \approx \frac{3}{N^2} \nonumber \\
< E_{F}(\symKet{1},\{1,1\}) & \approx & \frac{2\log_{2}N+
\log_{2}e}{N^2}
\end{eqnarray}
All measures monotonically decrease with $N$, but the distillable
entanglement decreases at least logarithmically faster than the
entanglement of formation. Similar statements are possible about
any $\symKet{m}$, with $N \gg m$.

\section{Entanglement in Symmetric Dynamically Generated States}
\label{Section::DynamicEntanglement}

Characterizing the reference states enabled us to quantitatively
identify the trade-off between the degree of entanglement and
robustness of the state to particle loss. This relationship can be
expressed as boundaries in the space expressed by the
entanglements of formation for the extremal splits. With this
relationship in hand, we are now able to address the question of
where various dynamically generated states lie with respect to all
accessible symmetric states.

For any given generation process, an important question involves
exactly how entanglement forms within an ensemble
\cite{horodecki1}. In this section, we characterize spin-squeezed
states, the most common experimental example of large scale
entanglement. It has been shown that spin-squeezing
(\refeqn{sscondition}) is a sufficient condition for an
$N$-particle system to be entangled \cite{sorensen1} and the
squeezing parameter also indicates in some sense the depth of
entanglement \cite{sorensen2}. It has also been demonstrated that
spin-squeezed systems contain significant pairwise entanglement
\cite{kitagawa2,wang}. However, little is known about the
entanglement of squeezed states across all size scales or how they
compare to the reference states from
\refsection{RepresentativeEntanglement}. Describing such states in
terms of entanglement measures is intrinsically important, but
also useful for understanding the more general class of symmetric
entangled states.  At the end of this section, we also briefly
discuss the problem of efficiently creating desirable states given
specified resources, allowable processes, and initially separable
states.

\subsection{Spin Squeezed States}

The collective angular momentum operators of any multipartite spin
state must satisfy the inequalities imposed by their commutation
relations. Let us assume without loss of generality that all
subsequent states satisfy $\langle J_{x} \rangle=\langle J_{y}
\rangle=0$ and $\langle J_{x}^{2} \rangle = \min_{\theta}(\langle
J_{\theta}^{2} \rangle)$ meaning $x$ is the direction of the
smallest variance perpendicular to the mean which points in the
$z$ direction. In this case, we use the uncertainty relationship
\begin{equation} \label{uncertainty}
    \langle J_{x}^{2} \rangle\langle J_{y}^{2}\rangle
        \geq \frac{\langle J_{z} \rangle^{2}}{4}
\end{equation}
The characteristic feature of spin-squeezed states is that
internal correlations between spins (i.e., entanglement) conspire
to reduce the noise in one angular momentum component ($x$) at the
expense of increasing the uncertainty in another ($y$). In
particular, spin squeezed states satisfy the inequality,
\begin{equation}
    \xi^2 \equiv \frac{N \langle J_{x}^{2} \rangle}
        {\langle J_{z}\rangle^{2}} < 1\\
\end{equation}
States with a minimal squeezing parameter, $\xi^2$, are useful for
reducing noise in many interferometric applications (e.g., atomic
clocks). Using Eq.\ (\ref{uncertainty}) and the fact that $\langle
J_{y}^{2} \rangle < J^{2}$, one can show that,
\begin{equation}
    \xi^2 > \frac{1}{N}
\end{equation}
where $1/N$ is the Heisenberg limit.

\subsubsection{Squeezing and Entanglement}

We choose to generate near optimally spin-squeezed states
$\ket{\Psi_{\xi}}$ by applying the counter-twisting Hamiltonian
$H_{ct}= (J_{+}^{2}-J_{-}^{2})/i$ to an initially polarized sample
$\ket{\Psi_{0}}=\symKet{0}$ (with $\xi^2=1$) for the length of
time $t_{N}$ needed to minimize $\xi^2$ \cite{kitagawa1}. This
process does not produce optimally squeezed states (see
\cite{sorensen2}), but in the large $N$ limit it creates states
which very nearly obtain the minimal value of $\xi^2$. The time it
takes to reach the minimum of $\xi^2 \propto 1/N$ for large $N$ is
$t_N \approx 0.2 \log_2(N)/N$ \cite{andre}. Henceforth, time is
scaled such that the optimal spin squeezing time $t_N=1$. We will
ignore the small difference between the achieved and optimum spin
squeezing so that we may examine the production of entanglement as
the state evolves in the most simple way. Interestingly, an
effective counter-twisting Hamiltonian can be experimentally
realized through the QND-detection and feedback rotation scheme of
\cite{thomsen}.

\begin{figure}
\centerline{\includegraphics[width=3in]{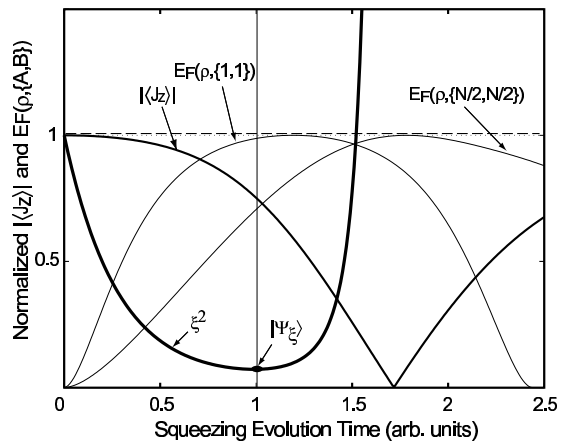}}
\caption{Spin squeezing evolution for a system of $N=50$ spin-1/2
particles evolving by the counter-twisting Hamiltonian as measured
by the squeezing parameter, $\xi^2$. The time is scaled such that
maximal spin squeezing occurs at $t=1$. The mean $J_z$ and the
entanglements of formation are all independently normalized by
their own maximum in the time period shown. Notice that the small
scale correlations, $E_F(\rho,\{1,1\})$, peak before the large
scale correlations, $E_F(\rho,\{N/2,N/2\})$, as the squeezing
evolves. \label{meansVtime}}
\end{figure}

Figure \ref{meansVtime} shows this evolution for a state with
$N=50$ spins.  The $x$ and $y$ means remain zero for the entire
evolution, while $\langle J_z \rangle$ decays from completely
polarized through zero.  For small numbers of spins, the state
will quickly re-cohere and become completely polarized
(separable). For large numbers of spins, the dynamics become
highly disordered after the mean decays through zero, and the
re-coherence time grows much longer.  After becoming maximally
squeezed, the internal entanglement continues to grow, but the
spin-squeezing rapidly gets worse because of the reduction in the
mean. The entanglement of formation for the largest and smallest
even splits [$E_F(\rho(t),\{N/2,N/2\})$ and
$E_F(\rho(t),\{1,1\})$] are also shown normalized by their own
initial local maximum.

\begin{figure}
\centerline{\includegraphics[width=3.2in]{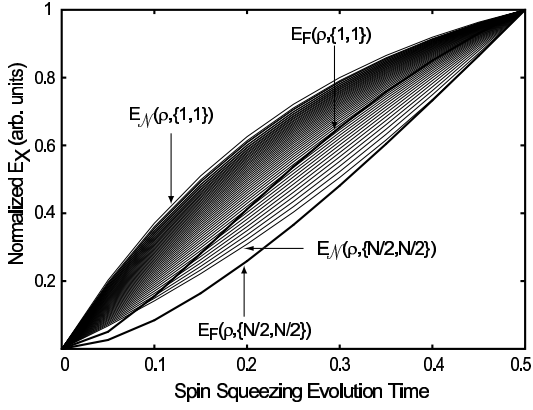}}
\caption{Entanglement measures for a system of $N=50$ spin-1/2
particles evolving under the influence of a counter-twisting
spin-squeezing Hamiltonian.  The time is scaled such that the
squeezing parameter achieves its minimum at $t=1$ (the small-time
evolution is depicted) and all entanglement measures are
independently normalized by their own maximum in the time period
shown. The entropy of formation, $E_F(\rho,\{A,B\})$, is shown for
the extremal bipartite splits, $\{1,1\}$ and $\{N/2, N/2\}$, while
the logarithmic negativity, $E_{\mathcal{N}}(\rho,\{A,B\})$, is
depicted for the partitions, $\{1,1\},
\{1,2\}$,$\{2,2\},\ldots,\{N/2-1,N/2\},\{N/2,N/2\}$.  It can be
seen that small-scale correlations tend to peak before their
large-scale counterparts; the entanglement measures are strictly
ordered according to the number of particles remaining
$N_r$.\label{negsVtime}}
\end{figure}

The small scale entanglement $\{1,1\}$ reaches its peak before the
large scale entanglement $\{N/2,N/2\}$ does. If we analyze the
relative rate of growth of the different scales of entanglement at
early times we see an intuitive ordering. Figure \ref{negsVtime}
shows the small-time logarithmic negativities (for all even
splits) and the entropy of formation (for the extremal splits)
normalized by their respective maxima over that interval. As the
state becomes squeezed, the $\{1,1\}$ correlations form first,
followed by the $\{1,2\}$, then the $\{2,2\}$, and so on up to
$\{N/2,N/2\}$. This observation suggests that small scale
correlations typically peak earlier than larger scale correlations
when evolving under quadratic Hamiltonians.

\begin{figure*}
\centerline{\includegraphics[width=6.2in]{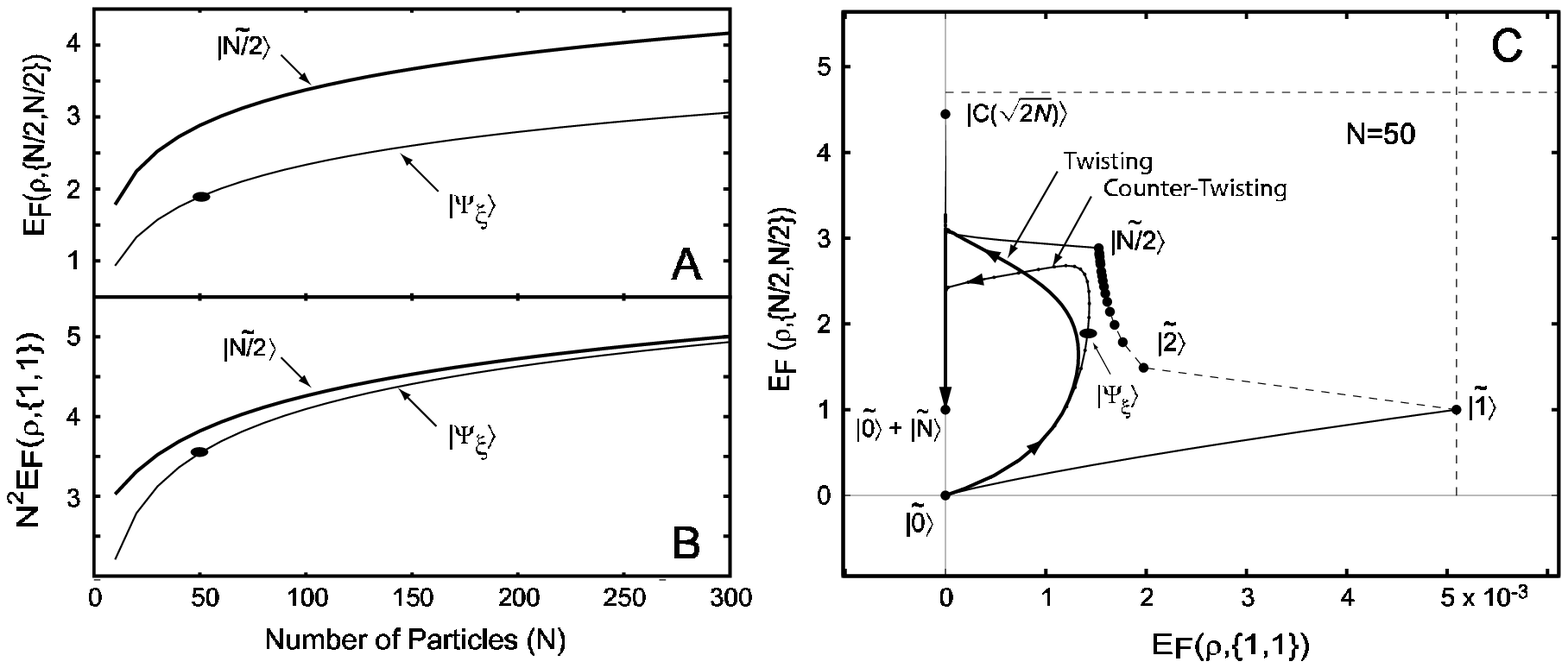}}
\caption{(A) A plot of the even split entanglement of formation
(and entropy), $E_{F}(\rho, \{N/2,N/2\})$, for a system of $N$
spin-1/2 particles evolved under a counter-twisting spin-squeezing
Hamiltonian.  The state $\ket{\Psi_{\xi}}$ minimizes the squeezing
parameter, $\xi^2$.  (B) A similar plot using the scaled
entanglement of formation, $N^2 E_F(\rho,\{1,1\})$, for a system
with all but two particles removed.  (C) The time evolution of
states evolving under both the counter-twisting Hamiltonian
($H_{ct}=(J_{+}^{2}-J_{-}^{2})/i$) and twisting Hamiltonian
($H_{t}=J_x^2$) in the space of extremal split entanglement.
\label{sscomposite}}
\end{figure*}

Another observation is that for small times, the state gets
progressively more entangled in the sense of majorization
\cite{nielsen}. In other words, the eigenvalues of
$\text{Tr}_{k}(\rho(t+dt))$ are more disordered than the
eigenvalues of $\text{Tr}_{k}(\rho(t))$ for all $k \leq N/2$ and
small $t$. Thus, despite certain ordering difficulties with
various entropies, the entanglement of any split is
\emph{strictly} increasing initially.

It is also important to quantitatively compare the entanglement
measures for spin-squeezed states and the symmetric reference
states.  Figure \ref{sscomposite}A shows the even split entropy
$E_F(\ket{\Psi_{\xi}},\{N/2,N/2\})$ of the optimally squeezed
state as a function of $N$. From a numerical fit, we find that
\begin{equation}
    E_F(\ket{\Psi_{\xi}},\{N/2,N/2\})\approx 0.46
        \log_{2}(N)-\log_{2}(e)
\end{equation}
For smaller scale entanglement, Figure \ref{sscomposite}B displays
the two-spin entropy $E_F(\ket{\Psi_{\xi}},\{1,1\})$.   The values
approach but never exceed the curve for $\symKet{N/2}$. Indeed, it
can be shown that in the large $N$ limit, the two-spin concurrence
scales identically for the two states: $\mathcal{C}(\symKet{N/2})
\approx \mathcal{C}(\ket{\Psi_{\xi}}) \approx 1/N$, thus the
entanglements of formation must also converge.

For a specified number of particles ($N=50$), the family of states
generated by applying the counter-twisting Hamiltonian to a
polarized sample are displayed in the
$[E_F(\rho(t),\{N/2,N/2\}),E_F(\rho(t),\{1,1\})]$ space of figure
\ref{sscomposite}C.  Again the small scale entanglement grows
faster than the large scale entanglement, but eventually decays to
zero as the large scale entanglement takes over.  The disordered
nature of the counter-twisting Hamiltonian dominates at long times
as the value of the large scale entanglement diffuses and the
small scale entanglement remains near zero.  In contrast, the
application of a twisting Hamiltonian $H_{t}=J_x^2$ (which, unlike
the counter-twisting Hamiltonian, creates squeezed states with a
rotating axis of squeezing) is seen to be much more periodic. The
states it generates are similar to the counter-twisting states
initially, but they eventually converge to the GHZ state then
return along the same trajectory.


The entropies of extremal splits ($\{1,1\}$ and $\{N/2,N/2\}$)
capture much of the character of a many-particle entangled state,
but there are of course a large number of other bipartite splits
to consider. The introduction of the information contained in all
other splits potentially brings up more interesting entanglement
characteristics. As in Figure \ref{negsVsplitref}, we can
efficiently calculate all even split bipartite logarithmic
negativities for large number states as they become spin squeezed.
The characteristic of early small scale entanglement being
transformed into subsequent large scale entanglement during the
course of the evolution is again apparent. Nonetheless, for this
particular case, the intermediate splits do not provide a
considerable amount of additional insight compared to that from
the extremal splits.

\subsubsection{Squeezing Under Particle Loss}

We now address how the spin-squeezing parameter behaves under
particle loss.  Given the expectation values of a set of operators
on a symmetric density matrix, it is simple to determine the
moments of the same state with a certain number of particles
removed. If $\rho_N$ is symmetric, so are all of its reduced
density matrices $\rho_{N_r}$ where $1\leq N_{r}\leq N$. Given
single particle operators $o_{i}$ we know that
\begin{equation}
\ptrace{N}(o_{i} \cdots o_{j}\rho_{N})=\ptrace{N-1}(o_{i} \cdots
o_{j}\rho_{N-1})
\end{equation}
assuming the indices of the operators are not the ones traced out.
With this observation and the fact that for symmetric states
\begin{eqnarray}
\langle J_{z}\rangle_{N}&=&N \text{Tr}_{N}(j_{z,i}\rho_{N})\\
\langle J_{z}\rangle_{N-1}&=&(N-1)
\text{Tr}_{N-1}(j_{z,i}\rho_{N-1})
\end{eqnarray}
we find
\begin{equation}
\langle J_{z}\rangle_{N-1}=\frac{N-1}{N}\langle J_{z}\rangle_{N}
\end{equation}
Similarly, it is easy to show
\begin{equation}
\langle J_{x}^{2}\rangle_{N-1}=\frac{N-2}{N}\langle
J_{x}^{2}\rangle_{N}+\frac{1}{4}
\end{equation}
Using these relations and taking the large $N$ limit we find that
the spin squeezing parameter of a state with $N_r$ spins remaining
($\xi_{N_r}^{2}$) is dependent on the initial squeezing parameter
($\xi_{N}^{2}$) and polarization of the state with all spins
remaining in the following way
\begin{equation} \label{Equation::sstrace}
\xi_{N_r}^{2}=\xi_{1}^{2}+(\xi_{N}^{2}-\xi_{1}^{2})\frac{N_r-1}{N-1}
\end{equation}
where $\xi_{1}^{2} \equiv N^{2}/(4\langle J_{z}\rangle_{N}^{2})$.

\begin{figure}
\centerline{\includegraphics[width=3in]{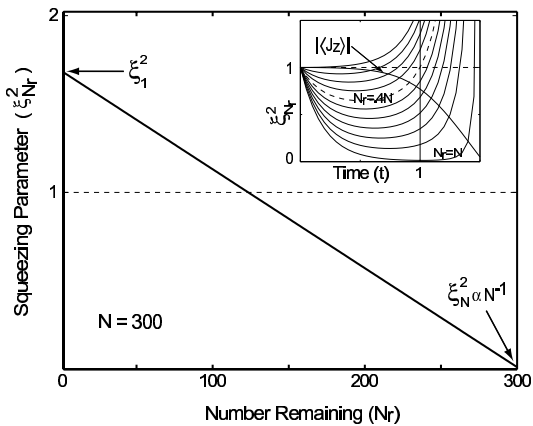}} \caption{Spin
squeezing reduction due to particle loss for a system initially
containing $N=300$ spin-1/2 particles.  The lower right corner of
the plot reflects the minimum value of the squeezing parameter,
$\xi^2$, achieved via a counter-twisting Hamiltonian.  As
particles are removed from the optimally squeezed system, the
value of $\xi^2_{N_r}$ moves up and left along the plotted line,
eventually crossing unity for finite number systems. The inset
shows the time evolution of $\xi^2$ for different numbers of
remaining particles, $N_r$. \label{ssVtrace}}
\end{figure}

The inset of Figure \ref{ssVtrace} shows the spin-squeezing
behavior for $N=300$ as a function of time. Considering only the
time when the state of all the spins is maximally squeezed
($t=1$), we plot the spin squeezing parameter as a function of
number remaining in Figure \ref{ssVtrace}, which behaves according
to \refeqn{sstrace}. For this finite number case, the
spin-squeezing is lost after some fraction of the spins are
removed.  As $N$ goes to infinity though, $\langle
J_{z}\rangle_{N}\rightarrow N/2$ ($\xi_{1}\rightarrow 1$) and
$\xi_{N}\rightarrow 0$ so all spins need to be removed for the
state to completely lose its spin squeezed character.

In a similar analysis, Simon and Kempe \cite{kempe} have shown
that spin-squeezed states remain squeezed until more than $29\%$
of the particles have \emph{depolarized}.  Thus spin-squeezed
states are robust to both particle loss and dephasing with constant $N$. (See
\cite{andre,kitagawa2} for a more complete treatment of how the
spin-squeezing parameter behaves under continuous generation and
decoherence.) However, robustness to particle loss and dephasing
do not necessarily imply each other because the GHZ states are
remarkably robust to local depolarization \cite{kempe}, but
obviously maximally fragile under particle loss. The complete
relationship between robustness to particle loss and dephasing
is an interesting direction for further research.

\subsection{Generating Entangled States}

Instead of characterizing what states a particular process
produces, consider the reverse problem of determining the process
necessary to generate a desired state from an initially separable
state.  The completely polarized initial state ($\symKet{0}$ or
$\symKet{N}$) is usually chosen both because it is completely
separable at all levels and it is easily prepared in the lab (e.g.
via optical pumping).

It can be shown that given such a state and access to Hamiltonians
of the form $J_x, J_y, J_z,$ and $J_x^2$, i.e., the generators are
all rotations plus a single non-linearity, one can produce
\emph{any} symmetric state by an algorithm that switches between
the Hamiltonians in time \cite{bacon}. Unfortunately, proving this
statement does not necessarily lead to the most efficient way to
create a particular state.  Knowing which states are prohibitively
expensive to produce is an important experimental question. An
interesting, but difficult, way to characterize a state is by
quantifying the resources needed to create that state given a
certain set of generators.  For example, one could define a cost
metric which is a function of how many times the Hamiltonians must
be switched and the length of time necessary to produce a
particular state.

Of course, all of these issues are context specific, but we can
summarize certain results.  Simply observing what the application
of a particular Hamiltonian produces is a first step.  The
counter-twisting Hamiltonian presented earlier produces optimal
squeezing but does not produce any recognizable reference state
(since the dynamics for large $N$ becomes highly disordered for
long times). A one-axis twisting Hamiltonian of $J_x^2$ produces
some squeezing which does not scale optimally \cite{kitagawa1}.
However, the time dependence of the entanglement produced by this
Hamiltonian is much more periodic and ordered than the
counter-twisting version. In fact, it produces the GHZ state
halfway through its period as is indicated in figure
\ref{sscomposite}C. M{\o}lmer and S{\o}rensen \cite{sorensen3}
have proposed a robust scheme for generating the GHZ state of many
hot ions taking advantage of this effect.

Unanyan \textit{et al.}\ have shown that by using adiabatic
passage and energy level navigation methods one can produce the
GHZ state and all $\symKet{m}$ states \cite{unanyan}.  However, it
remains unclear what the most efficient method is to generate
these states, or the bipartite entropy maximizing states presented
here, in the asymptotic limit of large $N$.


\section{Conclusion}

In this paper we analyzed the microscopic structure of
entanglement and its robustness to particle-loss for many-particle
symmetric states.  Our approach proceeded by comparing the
features of dynamically generated squeezed states to a collection
of symmetric representative states, including the GHZ and Dicke
states, as well as random states and a new family that we define.
In order to perform the analysis, we selected several bipartite
entanglement measures: the reduced entropy of entanglement, the
entanglement of formation, and the logarithmic negativity. By
computing these bipartite measures for all possible reductions and
partitions of the systems, we were able to construct a picture of
multi-scale entanglement.

Our analysis benefitted from simulations of many-particle systems.
The computational results helped to bolster physical insight and
provide a starting point for analytically treating the asymptotic
scaling of various entanglement measures.  In order to circumvent
the exponential scaling of the density matrix for arbitrary $N$
particle states, we restricted our analysis to the symmetric
subspace.  In \refsection{SymmetricSubspace} we developed
machinery for computing the above entanglement monotones for
symmetric states in a computationally efficient manner.  As a
result, our simulations were capable of handling systems with
$N\sim 10^3$ particles without making any dynamical
approximations.

In \refsection{RepresentativeEntanglement} we characterized the
entanglement of the representative states in detail, focusing on
the trade-off between those states that maximize the entanglement
measures and those that are robust under particle loss.  We also
analyzed several important ordering issues between the different
measures. A key point we stress is that fragility is not
necessarily a property of highly entangled states. With the
analysis in \refsection{RepresentativeEntanglement} it was
possible to address the evolution of microscopic entanglement in
dynamically generated spin-squeezed states.  Hopefully this work
helps clarify the otherwise vague statement that ``spin-squeezed
states are massively entangled.''

From this work, we anticipate several future directions.  First we
plan to consider less restrictive symmetry classes, particularly
the symmetry of \refeqn{doublesideP}. This symmetry is preserved
during the unconditional evolution of an ensemble under a uniform
symmetry-preserving Hamiltonian and local dephasing.  For certain
cases where the emission from the particles does not physically
distinguish different particles, the symmetry may also be
preserved under conditional evolution. In order to perform such an
analysis, it will be necessary to exploit the commutant algebra
and representation theory of the permutation group \cite{Sagan}.
Preliminary investigation suggests that it will be possible to
treat the full permutation group in a manner that scales
polynomially, rather than exponentially, with the number of
particles.

A more straightforward goal is to generalize the treatment of this
paper to particles with more than two levels. For example, we
would like to describe the entanglement within an ensemble of
Cesium atoms, where each atom can occupy the nine magnetic
sub-levels of the $F=4$ ground state.

Regarding dynamically generated states, it is possible to further
simplify the description of entanglement at small times.  For any
initially polarized state experiencing a quadratic Hamiltonian,
the state and relevant entanglement measures can be described in
terms of the Gaussian moments alone for short times. Deriving the
exact form of this low-dimensional parametrization of entanglement
measures is of experimental interest.

Finally, an important challenge is to develop techniques for
efficiently generating the reference states discussed in this
paper, including those with near maximal entanglement, such as the
comb states. Here, we hope to stress that theoretical treatments
of many-particle spin systems are most beneficial when they adopt
methods that can be experimentally implemented.

\begin{acknowledgements}
The authors acknowledge a number of important discussions with
Guifre Vidal, Dave Bacon, and Patrick Hayden. This work was
supported in part by the DoD Multidisciplinary University Research
Initiative (MURI) program administered by the Army Research Office
under grant No. DAAD19-00-1-0374 and the Caltech Institute for
Quantum Information sponsored by the National Science Foundation
under grant No. EIA-0086038. JKS acknowledges support from a Hertz
Foundation Fellowship.
\end{acknowledgements}

\appendix*
\section{Symmetric State Entropy Scaling}

\begin{theorem}
There exists a lower bound for the maximum attainable symmetric
state entropy that asymptotically scales as the maximum entropy
for states in $\mathbb{S}_{N/2} \otimes \mathbb{S}_{N/2}$,
\begin{eqnarray}
      \exists \ket{\Psi}  &\in&  \mathbb{S}_N , \delta>0, N^* > 0  : \forall N > N^*,\\
        && \log_2(N/2+1) - \eEntropy{\Psi}  <
        \delta    \nonumber
\end{eqnarray}
\end{theorem}

\begin{proof}
The proof proceeds by constructing a symmetric state whose even
split reduced entropy, $E(\ket{\Psi},\{N/2,N/2\})$ can be
expressed as the asymptotic series, $\log_2(N/2+1) + \delta$. In
order to obtain this series, we express the entropy in terms of
the Schmidt coefficients from \refeqn{SymmetricSchmidt} by
employing Stirling's formula. Computing the residuals that are
incurred by Stirling's approximation, we obtain a bound for
$\delta$ and demonstrate that it is asymptotically constant in
$N$, i.e., $\delta \sim O(1)$.

Consider the family of $\cState{s}$ states, defined in
\refsection{CStateDefinition}, whose Schmidt decomposition,
according to \refeqn{SymmetricSchmidt}, is given by,
\begin{equation}
    \cState{s} = A \sum_{m=-N/s}^{N/s}
        \sum_{i=0}^{N-k} \sum_{j=0}^k
        \frac{\C{N}{m}\delta_{\frac{N}{2}+ms,i+j}}{\C{N-k}{i}\C{k}{j}} \symKet{i}_{N-k}
        \symKet{j}_k
\end{equation}
where $A=\sqrt{2s/N}$. We wish to choose the value of $s$ in
$\cState{s}$ such that the matrix, $\mathbf{c}$, becomes
block-diagonal in the large $N$ limit, which will happen provided
that the $\symKet{m}$ contributing to $\cState{s}$ are
sufficiently separated. For an orthogonal Schmidt matrix,
\refeqn{SchmidtSVD} can be solved in closed form and the total
entropy is a weighted sum of the entropies contributed by each
participating, $\symKet{m}$.

The required separation between nonzero $\symKet{m}$  in
$\cState{s}$ as well as their contributing entropies, can be found
by considering $c_m(i)$, the $c_{ij}$ matrix elements as a
function of $i$ for a given value of $m$. This leads to the
distribution,
\begin{equation}
    c_m(i) = \sqrt{\frac{\choose{N-k}{i}
        \choose{k}{m-i}}{\choose{N}{m}}}
\end{equation}
which can be approximated using Stirling's formula,
\begin{equation}
    \log n! = n \log n - n + \sqrt{2\pi n}
        + \frac{1}{12n} + O(n^{-2})
\end{equation}
to obtain the expression,
\begin{eqnarray}
    c_m^2(i) & = & 2^{-(m+\frac{1}{2})}
    \exp\left[\frac{i^2-i+m+m^2}{12 i^2m-12 i m^2}\right] \times
    \label{Equation::EntropySeries}\\
    & & i^{-(i+\frac{1}{2})}
        m^{m+\frac{1}{2}} (m-i)^{i-m-\frac{1}{2}}
        \pi^{-\frac{1}{2}} + O(m^{-2}) \nonumber
\end{eqnarray}
for $c^2_m(i)$ as $N \to \infty$.  The residual terms are of order
$m^{-2}$.

Unfortunately, it is not known how to evaluate the entropy of
\refeqn{EntropySeries} because the discrete sum,
\begin{equation} \label{Equation::DiscreteEntropy}
    S(m) = -\sum_{i=0}^{\infty} c_m^2(i) \log_2 c_m^2(i)
\end{equation}
cannot be expressed in closed form. However, the moments of
$c_m(i)$ can be computed analytically,
\begin{eqnarray}
    \bar{i} & = & \langle i \rangle = \frac{m}{2}
    \label{Equation::CMean} \\
    \sigma^2 & = & \langle i^2 \rangle -
            \langle i \rangle^2 = \frac{1}{12}m(m+2)
    \label{Equation::CVariance}
\end{eqnarray}
and it can be seen that all higher cumulants vanish in the large
$N$ limit.

The entropy contributed by each $\symKet{m}$ in $\cState{s}$ when
$m$ is large can be computed by approximating the $c^2_m(i)$ as
Gaussian,
\begin{equation} \label{Equation::CGaussian}
    c^2_m(i) = \sqrt{\frac{1}{m\pi}}
        \exp\left[\frac{-1}{4m}-\frac{\sqrt{3}
            (i-m/2)^2}{\sqrt{m(m+2)}}\right] + O(m^{-2})
\end{equation}
and transforming the summation in \refeqn{DiscreteEntropy} into an
integral,
\begin{equation}
    S_m - \epsilon = -\int_0^\infty
        c_m^2(i) \log_2 c_m^2(i) \,di
        + \int O(m^{-2}) \,di
\end{equation}
where the error term, $\epsilon$, must be obtained using the
Abel-Plana procedure \cite{hardy} for computing the difference
between a discrete sum and its corresponding integral.  An upper
bound for the integral over the residual $O(m^{-2})$ terms can be
found to converge to, $\varepsilon = \sqrt{2 \pi} m^{-2}
\exp(-m^2)/128$, by considering the asymptotic series of the
$\Gamma$ function \cite{abramowitz}.

The resulting entropy, in the large $m$ limit, with the Abel-Plana
corrections and the error bounds from higher order Stirling terms
can be shown to have the form,
\begin{eqnarray}
    S_m & = & \frac{e^{-1/4m}m^{1/4}(2+m)^{3/4}}{
    3^{1/4} \sqrt{2m(2+m)}\log 2} \\
    & & +
    \vartheta_3(0,m^{-\frac{1}{2}} e^{-1/4m}) - \frac{1}{2}
    \vartheta_4(0,1/ \sqrt{\pi}) + \varepsilon \nonumber
\end{eqnarray}
where $\vartheta_n(u,v)$ is the elliptic theta function of order
$n$ \cite{abramowitz}.  Computing the entropy of the full state
requires evaluating the discrete sum over $m$.  We chose the comb
spacing to be $s=\sqrt{2 N}$ based on \refeqn{CVariance}.
Performing the final sum leads to,
\begin{equation}
    E(\ket{\Psi},\{N/2,N/2\}) = (1+\frac{1}{2 \sqrt{N}}) \log_2 N + O(N
    e^{-\sqrt{N}}) + \varepsilon
\end{equation}
where the first residual term reflects the finite overlap of the
$c_m^2(i)$, i.e. corrections that arise because $\mathbf{c}$ is
not perfectly block-diagonal.

Finally it is possible to express the reduced entropy as an
asymptotic series,
\begin{equation}
    E(\ket{\Psi},\{N/2,N/2\}) \rightarrow \log_2(N/2+1) -  O(1)
\end{equation}
which completes the proof.

\end{proof}

\end{document}